\documentclass[journal,onecolumn,final,twoside,letterpaper,11pt]{IEEEtran}
\usepackage{subfigure,cite,graphicx,amsmath,amssymb,eufrak,mathrsfs,epsf,epsfig}
\begin{document}

\title{Capacity of a Class of Linear Binary Field Multi-source Relay Networks}
\author{Sang-Woon Jeon, \emph{Student Member}, \emph{IEEE} and Sae-Young Chung, \emph{Senior Member}, \emph{IEEE}\\
\thanks{S.-W. Jeon and S.-Y. Chung are with the Department of EE, KAIST, Daejeon, South Korea (e-mail: swjeon@kaist.ac.kr; sychung@ee.kaist.ac.kr).}
\thanks{The material in this paper was presented in part at the Information Theory and Applications Workshop, University of California San Diego, La Jolla, CA, February 2009, and at the IEEE International Symposium on Information Theory (ISIT), Seoul, Korea, June/July 2009.}
}

\maketitle


\newtheorem{definition}{Definition}
\newtheorem{theorem}{Theorem}
\newtheorem{lemma}{Lemma}
\newtheorem{example}{Example}
\newtheorem{corollary}{Corollary}
\newtheorem{proposition}{Proposition}
\newtheorem{conjecture}{Conjecture}
\newtheorem{remark}{Remark}

\def \diag{\operatornamewithlimits{diag}}
\def \min{\operatornamewithlimits{min}}
\def \max{\operatornamewithlimits{max}}
\def \log{\operatorname{log}}
\def \max{\operatorname{max}}
\def \rank{\operatorname{rank}}
\def \out{\operatorname{out}}
\def \exp{\operatorname{exp}}
\def \arg{\operatorname{arg}}
\def \E{\operatorname{E}}
\def \tr{\operatorname{tr}}
\def \SNR{\operatorname{SNR}}
\def \SINR{\operatorname{SINR}}
\def \dB{\operatorname{dB}}
\def \ln{\operatorname{ln}}
\def \th{\operatorname{th}}

\begin{abstract}
Characterizing the capacity region of multi-source wireless relay
networks is one of the fundamental issues in network information
theory. The problem is, however, quite challenging due to inter-user interference when
there exist multiple source--destination (S--D) pairs in the
network. By focusing on a special class of networks, we show that
the capacity can be found. Namely, we study a layered linear
binary field network with time-varying channels, which is a simplified model reflecting
broadcast, interference, and fading natures of wireless
communications. We observe that fading can play an important role in
mitigating inter-user interference effectively for both single-hop
and multi-hop networks. We propose new encoding and
relaying schemes with randomized channel pairing, which exploit such
channel variations, and derive their achievable rates. By comparing
them with the cut-set upper bound, the capacity region of
single-hop networks and the sum capacity of multi-hop networks can
be characterized for some classes of channel distributions and
network topologies. For these classes, we show that the capacity
region or sum capacity can be interpreted as the max-flow min-cut
theorem.
\end{abstract}

\section{Introduction} \label{sec:intro}
Capacity characterization of general wireless relay networks is a
fundamental problem in network information theory. However, the
capacity is not fully characterized even for the simplest network
consisting of single source, single relay, and single destination
\cite{Cover:79}. In wireless environments, a transmit signal will be
heard by multiple nodes, which we call the \emph{broadcast} nature
of wireless communications, and a receiver will receive the
superposition of simultaneously transmitted signals from multiple
nodes, which we call the \emph{interference} nature of wireless
communications. Furthermore wireless channels may be time-varying due
to \emph{fading}, and there is noise at each receiver. Considering
all these makes the problem vary hard.

Hence, one of the promising approaches is to study simplified relay
networks, whose results can provide insights towards exact or
approximate capacity characterization for more general wireless
relay networks. Let us first look at some cases for which the capacity is
known. For wireline relay networks, routing is enough to achieve the
unicast capacity \cite{Ford:62}. On the other hand, routing alone
cannot achieve the multicast capacity and network coding has been
shown to be optimal in this case \cite{Ahlswede:00, Li:03,
Koetter:03, Ho:06}. For deterministic relay networks with no
interference, the unicast capacity has been characterized in
\cite{ArefPD:80} and the extension to the multicast case has been
studied in \cite{Niranjan:06}. The multicast capacity of erasure
networks with no interference has been also characterized in
\cite{Amir:06}. When there is no broadcast, the unicast capacity of
erasure networks has been characterized in \cite{Smith:07}, which is
the dual network studied in \cite{Amir:06}. For all these mentioned
networks, the unicast or multicast capacity can be interpreted as
the \emph{max-flow min-cut theorem}.

Notice that although such orthogonal transmission or reception is
possible in practice by using time, frequency, or code-division
techniques, it is suboptimal in general. Therefore,
simplification of wireless relay networks while preserving both
broadcast and interference natures is crucially important to capture
the essence of wireless communications. One of the simplest models that
successfully reflect both broadcast and interference natures is a linear finite field relay network \cite{Ray:03, Bhadra:06,
AvestimehrDiggaviTse:07}, where a node transmits an element in the
finite field and receives the sum of transmit signals in the same
finite field. Recently, the work in \cite{AvestimehrDiggaviTse:07}
has shown that the max-flow min-cut theorem also holds for
deterministic linear finite field relay networks.
After the capacity characterization of linear finite field relay
networks, the approximate capacity of Gaussian relay networks has
been characterized within a constant number of bits/s/Hz using the
quantize-random-map-and-forward by the same authors
\cite{AvestimehrDiggaviTse:08}.

In spite of the surging importance of multi-source relay networks,
capacity characterization is much more challenging if there exist
multiple source--destination (S--D) pairs in a network. Even for
linear finite field relay networks, the extension of the results in
\cite{AvestimehrDiggaviTse:07} to the multi-source does not seem to
be straightforward. Notice that the main difficulty arises from the
fact that the transmission of other sessions acts as
\emph{inter-user interference} and, as a result, the cut-set upper
bound is not tight in general.
Due to these difficulties, the existing capacity or approximate
capacity results are limited in specific network topologies such as
two-user interference channel \cite{Etkin:08,Bresler:08},
many-to-one and one-to-many interference channel \cite{Bresler:07},
two-way channel \cite{Avestimehr:08, Nam:00}, two-user two-hop relay
network \cite{Mohajer:08, Mohajer:09}, and double Z-channel
\cite{Aggarwal:09}. Therefore, one of the basic questions is whether
we can characterize the capacity or approximate capacity for more
general network topologies or other classes of relay networks.

In this paper, we study a layered \emph{multi-source linear
binary field relay network with time-varying channels}, which
captures three key characteristics of wireless environment, i.e.,
broadcast, interference, and fading. Note that a random coding
strategy, which is still optimal in fading single-source networks
\cite{Haddad:09, Lim:09}, does not work anymore for our network model due to the
inter-user interference. As mentioned before, a fundamental issue in
multi-source networks is how to manage inter-user interference properly. We
observe that fading can play an important role in mitigating such
interference efficiently, which leads to the capacity
characterization for certain classes of networks. More specifically,
for single-hop networks, inter-user interference can be removed
completely at each destination by using two particular channel
instances jointly. For multi-hop networks, by using a series of
particular channel instances over multiple hops, each destination
can also decode its message without interference.

\begin{figure}[t!]
  \begin{center}
  \scalebox{1.1}{\includegraphics{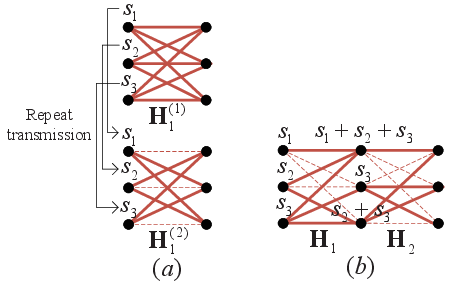}}
  \caption{Interference mitigation for the single-hop network (a) and for the two-hop network (b), where the solid lines and the dashed lines denote the corresponding channels are ones and zeros, respectively.}
  \label{FIG:interference_alignment_mitigation}
  \end{center}
\end{figure}

As an example, consider the three-user linear binary field relay
network in Fig. \ref{FIG:interference_alignment_mitigation}, where
$s_k\in\mathbb{F}_2$ denotes the information bit of the $k$-th
source and the symbol in each node denotes the transmit signal of
that node.
For single-hop networks, as shown in Fig.
\ref{FIG:interference_alignment_mitigation}. (a), by transmitting the
same bit twice at each source through $\mathbf{H}^{(1)}_1$ and
$\mathbf{H}^{(2)}_1$ such that
$\mathbf{H}^{(1)}_1+\mathbf{H}^{(2)}_1=\mathbf{I}$, each destination
can cancel interference by adding the two received signals, where
$\mathbf{H}^{(1)}_1$ and $\mathbf{H}^{(2)}_1$ denote the two
different channel instances of the first hop and  $\mathbf{I}$
denotes the identity matrix. Related works dealing with the
inseparability of parallel interference channels can be found in
\cite{CadambeJafar:08, Sankar:08, JeonITA:09, Nazer:09} and the
references therein.
The idea of opportunistically pairing two
channel instances, i.e.,
$\mathbf{H}^{(1)}_1+\mathbf{H}^{(2)}_1=\mathbf{I}$, also appeared in
\cite{JeonITA:09, Nazer:09}. This can be considered as a different
and simpler way of doing interference alignment
\cite{Viveck1:08,Viveck2:09}. For two-hop networks, as shown Fig.
\ref{FIG:interference_alignment_mitigation}. (b), we notice that
each destination can receive the information bit without
interference if $\mathbf{H}_2\mathbf{H}_1=\mathbf{I}$, where
$\mathbf{H}_1$ and $\mathbf{H}_2$ denote the channel instances of
the first and second hop, respectively.
In general, the interference-free communication is possible for
$M$-hop networks, $M\geq 2$, by opportunistically pairing the series
of channel instances from $\mathbf{H}_1$ to $\mathbf{H}_M$ such that
$\mathbf{H}_M\mathbf{H}_{M-1}\cdots \mathbf{H}_1=\mathbf{I}$, where
$\mathbf{H}_m$ denotes the channel instance of the $m$-th hop.


Based on these key observations, we propose encoding
and relaying schemes which make such opportunistic pairing of
channel instances possible. By comparing their achievable rate
regions with the cut-set upper bound, we characterize the capacity
region of single-hop networks and the sum capacity of multi-hop
networks for some classes of  network topologies and channel
distributions.

This paper is organized as follows. In Section \ref{sec:sys_model},
we define the network model and state the multi-source relay problem
and the notations used in the paper. In Section \ref{sec:converse},
we derive the general cut-set upper bound, which will be used to
prove the converses in Section \ref{sec:achievability}. In Section
\ref{sec:achievability}, new encoding and relaying schemes
are proposed to mitigate inter-user interference, which
characterizes the capacity region or sum capacity for certain
classes of networks. We conclude this paper in Section
\ref{sec:conclusion} and refer the proofs of the lemmas to
Appendices I and II.

\section{System Model} \label{sec:sys_model}
In this section, we first explain the underlying network model and
then define the achievable rate region and the notations used in the
paper. Throughout the paper, $\mathbf{A}$ and $\mathbf{a}$ denote a
matrix and a vector, respectively. The symbol $\mathcal{A}$ denotes
a set and $|\mathcal{A}|$ denotes the cardinality of $\mathcal{A}$.

\begin{figure}[t!]
  \begin{center}
  \scalebox{0.9}{\includegraphics{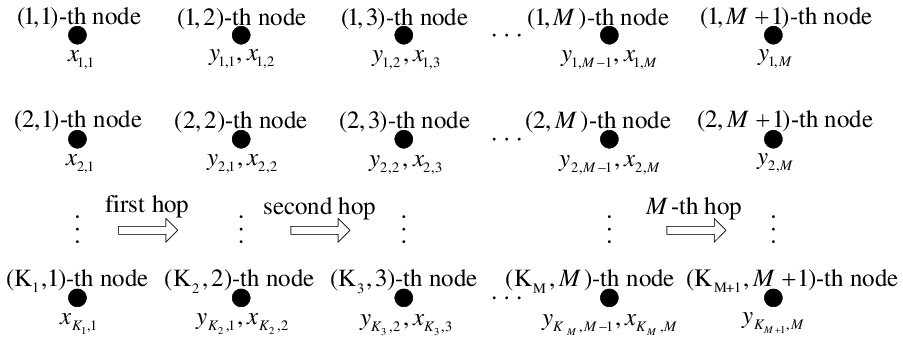}}
  \caption{Layered multi-source relay network.}
  \label{FIG:Kuser_Mhop}
  \end{center}
\end{figure}

\subsection{Linear Binary Field Relay Networks}
We study a layered network in Fig. \ref{FIG:Kuser_Mhop} that
consists of $M+1$ layers having $K_m$ nodes at the $m$-th layer,
where $m\in\{1,\cdots,M+1\}$. Let us denote
$K_{\operatorname{max}}=\max_m\{K_m\}$ and
$K_{\operatorname{min}}=\min_m\{K_m\}$. The $(k,m)$-th node refers
to the $k$-th node at the $m$-th layer. Then $K=K_1=K_{M+1}$ is the number
of S--D pairs and the $(k,1)$-th node and
the $(k,M+1)$-th node are the source and the destination of the
$k$-th S--D pair, respectively. Notice that if $M=1$, the network becomes a $K$-user
interference channel.

Consider the $m$-th hop transmission. The $(i,m)$-th node and the
$(j,m+1)$-th node become the $i$-th transmitter (Tx) and the $j$-th
receiver (Rx) of the $m$-th hop, respectively, where
$i\in\{1,\cdots,K_m\}$ and $j\in\{1,\cdots,K_{m+1}\}$. Let
$x_{i,m}[t]\in \mathbb{F}_2$ denote the transmit signal of the
$(i,m)$-th node at time $t$ and $y_{j,m}[t]\in \mathbb{F}_2$ denote
the received signal of the $(j,m+1)$-th node at time $t$. Let
$h_{j,i,m}[t]\in \mathbb{F}_2$ be the channel from the $(i,m)$-th
node to the $(j,m+1)$-th node at time $t$. The relation between the
transmit and received signals is given by
\begin{equation}
y_{j,m}[t]=\sum_{i=1}^{K_m} h_{j,i,m}[t]x_{i,m}[t],
\end{equation}
where all operations are performed over $\mathbb{F}_2$\footnote{We
focus on the binary field $\mathbb{F}_2$ in this paper, but some
results can be directly extended to $\mathbb{F}_q$ (see Remarks
\ref{RE:q_ary_single} and \ref{RE:q_ary_multi}).}. We assume
time-varying channels such that
\begin{equation}
\Pr(h_{j,i,m}[t]=1)=p_{j,i,m}
\end{equation}
and $h_{j,i,m}[t]$ are independent of each other for different $i$,
$j$, $m$, and $t$. This assumption can be generalized to block
fading with coherence time of $T$ symbols, where $T\gg 1$ such that there is enough time for CSI to be spread to relevant nodes.
We assume $T=1$ for notational simplicity since our result does not explicitly depend on $T$ as long as it is big enough such that CSI is available at all relevant nodes. Let
$\mathbf{x}_m[t]$ and $\mathbf{y}_m[t]$ be the $K_m\times1$ transmit
signal vector and $K_{m+1}\times1$ received signal vector of the
$m$-th hop, respectively, where
$\mathbf{x}_m[t]=\left[x_{1,m}[t],\cdots,x_{K_m,m}[t]\right]^T$,
$\mathbf{y}_m[t]=\left[y_{1,m}[t],\cdots,y_{K_{m+1},m}[t]\right]^T$.
Then the transmission of the $m$-th hop can be represented as

\begin{equation}
\mathbf{y}_m[t]=\mathbf{H}_m[t]\mathbf{x}_m[t],
\end{equation}
where $\mathbf{H}_m[t]$ is the $K_{m+1}\times K_m$ channel matrix of
the $m$-th hop having $h_{j,i,m}[t]$ as the $(j,i)$-th element. We
assume that both Txs and Rxs of the $m$-th hop causally know the
global channel state information (CSI) up to the $m$-th hop.
That
is, at time $t_0$, the nodes in the $m$-th layer know
$\{\mathbf{H}_1[t],\cdots,\mathbf{H}_m[t]\}_{t=1}^{t_0}$ if $m\leq
M$ and   $\{\mathbf{H}_1[t],\cdots,\mathbf{H}_M[t]\}_{t=1}^{t_0}$ if
$m= M+1$.

For a broad class of networks, if the channel dimension of a certain
hop is smaller than those of the other hops, then the average
channel rank of the hop is likely to be less than those of the
other hops. The following definition formally states this class of
networks.
\begin{definition} \label{Def:miminal_dim}
Let
$m_0=\arg\min_{m\in\{1,\cdots,M\}}\mathbb{E}(\rank(\mathbf{H}_m[1]))$\footnote{Notice
that $\mathbb{E}(\rank(\mathbf{H}_m[t]))$ is the same for all $t$.}.
A linear binary relay network is said to have a
\emph{minimum-dimensional bottleneck-hop} $m_0$ if $K_m\geq K_{m_0}$
and $K_{m+1}\geq K_{m_0+1}$ or $K_m\geq K_{m_0+1}$ and $K_{m+1}\geq
K_{m_0}$ for all $m\in\{1,\cdots,M\}$.
\end{definition}
In this paper, we will study the class of networks satisfying
Definition \ref{Def:miminal_dim}. Notice that any networks with
$K_m=K$ for all $m\in\{1,\cdots,M+1\}$ or any one-hop or two-hop
networks are included in this class of networks regardless of
channel distributions.

\subsection{Problem Statement}
Based on the previous network model, we define a set of length-$n$
block codes. Let $W_k$ be the message of the $k$-th source uniformly
distributed over $\{1,2,\cdots,2^{nR_k}\}$, where $R_k$ is the rate
of the $k$-th source. For simplicity, we assume $nR_k$ is an
integer. Then a $\left(2^{nR_1},\cdots,2^{nR_K};n\right)$ code
consists of the following encoding, relaying, and decoding
functions.

\begin{itemize}
\item (Encoding)

For $k\in\{1,\cdots,K\}$, the set of encoding functions of the
$k$-th source is given by
$\{f_{k,1,t}\}_{t=1}^n:\{1,\cdots,2^{nR_k}\}\to \mathbb{F}_2^n$ such
that
\begin{equation}
x_{k,1}[t]=f_{k,1,t}(W_k) \mbox{ for } t\in\{1,\cdots,n\}.
\end{equation}
\item (Relaying)

For $m\in\{2,\cdots,M\}$ and $k\in\{1,\cdots,K_m\}$, the set of
relaying functions of the $(k,m)$-th node is given by
$\{f_{k,m,t}\}_{t=1}^n:\mathbb{F}_2^n\to \mathbb{F}_2^n$ such that
\begin{equation}
x_{k,m}[t]=f_{k,m,t}\left(y_{k,m-1}[1],\cdots,
y_{k,m-1}[t-1]\right)\mbox{ for } t\in\{1,\cdots,n\}.
\end{equation}
\item (Decoding)

For $k\in\{1,\cdots,K\}$, the decoding function of the $k$-th
destination is given by $g_k:\mathbb{F}_2^n\to\{1,\cdots,2^{nR_k}\}$
such that
\begin{equation}
\hat{W}_k=g_k\left(y_{k,M}[1],\cdots,y_{k,M}[n]\right).
\end{equation}
\end{itemize}

If $M=1$, the sources transmit directly to the destinations without
relays. The probability of error at the $k$-th destination is given
by $P^{(n)}_{e,k}=\Pr(\hat{W}_k\neq W_k)$. A set of rates
$\left(R_1,\cdots,R_K\right)$ is said to be \emph{achievable} if
there exists a sequence of $(2^{nR_1},\cdots,2^{nR_K};n)$ codes with
$P^{(n)}_{e,k}\to 0$ as $n\to\infty$ for all $k\in\{1,\cdots,K\}$.
Then the achievable sum rate is simply given by
$R_{\operatorname{sum}}=\sum_{k=1}^{K}R_k$. The capacity region is
the closure of all achievable $(R_1,\cdots,R_K)$ and the sum
capacity is the supremum of all achievable sum rates.

\subsection{Notations}
In this subsection, we introduce the notations for directed graphs
and define sets of channel instances and sets of nodes.

\subsubsection{Notations for directed graphs}
The considered network can be represented as a directed graph
$\mathcal{G}=(\mathcal{V},\mathcal{E})$ consisting of a vertex set
$\mathcal{V}$ and a directed edge set $\mathcal{E}$. Let $v_{k,m}$
denote the $(k,m)$-th node and
$\mathcal{V}_m=\{v_{k,m}\}_{k=1}^{K_m}$ denote the set of nodes in
the $m$-th layer. Then $\mathcal{V}$ is given by
$\cup_{m\in\{1,\cdots,M+1\}}\mathcal{V}_m$. The sets of sources and
destinations are given by $\mathcal{S}=\mathcal{V}_1$ and
$\mathcal{D}=\mathcal{V}_{M+1}$, respectively.

There exists a directed edge $(v_{i,m},v_{j,m+1})$ from $v_{i,m}$ to
$v_{j,m+1}$ if $p_{j,i,m}>0$. For $\mathcal{V}'\subseteq
\mathcal{V}$ and $\mathcal{V}''\subseteq \mathcal{V}$, define
$\mathcal{E}(\mathcal{V}',\mathcal{V}'')$ as the set of edges going
from $\mathcal{V}'$ to $\mathcal{V}''$ given by
$\{(v',v'')|v'\in\mathcal{V}',v''\in\mathcal{V}'',
(v',v'')\in\mathcal{E}\}$. We say node $v''$ is \emph{reachable} from
node $v'$ if there exists a series of edges from $v'$ to $v''$,
where we assume $v'$ is always reachable from $v'$ itself. We further
define $v''$ is reachable under $\mathcal{V}'$ from $v'$ if there
exists a series of edges in $\mathcal{E}(\mathcal{V}',\mathcal{V}')$
from $v'$ to $v''$. We define cut $\Omega\subseteq \mathcal{V}$ as a
subset of nodes such that at least one source is in $\Omega$ and at
least one corresponding destination is in $\Omega^c$. We define the
following sets related to $\Omega$:
\begin{eqnarray}
\mathcal{K}_{\Omega}\!\!\!\!\!\!\!\!\!&&=\{k|v_{k,1}\in\Omega,v_{k,M+1}\in\Omega^c,k\in\{1,\cdots,K\}\},\nonumber\\
\mathcal{D}_{\Omega}\!\!\!\!\!\!\!\!\!&&=\{v_{k,M+1}|k\in\mathcal{K}_{\Omega}\},\nonumber\\
\mathcal{S}_{\Omega}\!\!\!\!\!\!\!\!\!&&=\{v_{k,1}|k\in\mathcal{K}_{\Omega}\},\nonumber\\
\Omega_D\!\!\!\!\!\!\!\!\!&&=\{v|\mathcal{E}(\Omega,\{v\})\neq\phi, \mbox{ at least one of the destinations in $\mathcal{D}_{\Omega}$}\nonumber\\
&&{~~~~}\mbox{is reachable under $\Omega^c$ from $v$}, v\in\Omega^c\},\nonumber\\
\Omega'\!\!\!\!\!\!\!\!\!&&=\{v|\mbox{$v\in\Omega$ is reachable from at least one of the sources in $\mathcal{S}_{\Omega}$}\},\nonumber\\
\Omega_S\!\!\!\!\!\!\!\!\!&&=\{v|\mathcal{E}(\{v\},\Omega_D)\neq
\phi,v\in\Omega'\}.
\end{eqnarray}

Let $\mathcal{X}_{\mathcal{V}'}[t]$ and
$\mathcal{Y}_{\mathcal{V}'}[t]$ denote the sets of transmit and
received signals of the nodes in $\mathcal{V}'$ at time $t$,
respectively. Let $\mathbf{H}_{\mathcal{V}',\mathcal{V}''}[t]$ be
the $|\mathcal{V}''|\times|\mathcal{V}'|$ channel matrix at time $t$
from the nodes in $\mathcal{V}'$ to the nodes in $\mathcal{V}''$.
Hence
$\mathbf{H}_{\mathcal{V}_m,\mathcal{V}_{m+1}}[t]=\mathbf{H}_m[t]$.
For notational simplicity, we use $\mathbf{H}_{\Omega}[t]$ to denote
$\mathbf{H}_{\Omega_S,\Omega_D}[t]$ in this paper.

\subsubsection{Sets of channel instances and nodes}
For $\bar{\mathcal{V}}'\subseteq\mathcal{V}'$,
$\bar{\mathcal{V}}''\subseteq\mathcal{V}''$, and
$\mathbf{G}\in\mathbb{F}_2^{|\bar{\mathcal{V}}''|\times
|\bar{\mathcal{V}}'|}$, we define the following sets of channel
instances.
\begin{eqnarray}
\mathcal{H}_{\mathcal{V}',\mathcal{V}''}\left(\mathbf{G}, \bar{\mathcal{V}}',\bar{\mathcal{V}}''\right)\!\!\!\!\!\!\!\!\!&&=\big\{\mathbf{H}_{\mathcal{V}',\mathcal{V}''}[1]\big|\mathbf{H}_{\bar{\mathcal{V}}',\bar{\mathcal{V}}''}[1]=\mathbf{G},\mathbf{H}_{\mathcal{V}',\mathcal{V}''}[1]\in \mathbb{F}_2^{|\mathcal{V}''|\times|\mathcal{V}'|}\big\},\nonumber\\
\mathcal{H}^F_{\mathcal{V}',\mathcal{V}''}\left(\mathbf{G}, \bar{\mathcal{V}}',\bar{\mathcal{V}}''\right)\!\!\!\!\!\!\!\!\!&&=\big\{\mathbf{H}_{\mathcal{V}',\mathcal{V}''}[1]\big|\operatorname{rank}(\mathbf{H}_{\mathcal{V}',\mathcal{V}''}[1])=\operatorname{rank}(\mathbf{G}),\mathbf{H}_{\bar{\mathcal{V}}',\bar{\mathcal{V}}''}[1]=\mathbf{G},\nonumber\\
&&{~~~~~}\mathbf{H}_{\mathcal{V}',\mathcal{V}''}[1]\in
\mathbb{F}_2^{|\mathcal{V}''|\times|\mathcal{V}'|}\big\}.
\end{eqnarray}
Note that $\mathcal{H}_{\mathcal{V}',\mathcal{V}''}\left(\mathbf{G},
\bar{\mathcal{V}}',\bar{\mathcal{V}}''\right)$ is the set of all
$\mathbf{H}_{\mathcal{V}',\mathcal{V}''}[1]\in \mathbb{F}_2^{|\mathcal{V}''|\times|\mathcal{V}'|}$ that contain $\mathbf{G}$ in
$\mathbf{H}_{\bar{\mathcal{V}}',\bar{\mathcal{V}}''}[1]$.
Similarly,
$\mathcal{H}^F_{\mathcal{V}',\mathcal{V}''}\left(\mathbf{G},
\bar{\mathcal{V}}',\bar{\mathcal{V}}''\right)$ is the set of all
$\mathbf{H}_{\mathcal{V}',\mathcal{V}''}[1]\in \mathbb{F}_2^{|\mathcal{V}''|\times|\mathcal{V}'|}$ that have the same rank as $\mathbf{G}$ and contain
$\mathbf{G}$ in
$\mathbf{H}_{\bar{\mathcal{V}}',\bar{\mathcal{V}}''}[1]$.

We further define the following sets of nodes. For positive integers
$a\leq |\mathcal{V}'|$ and $b\leq |\mathcal{V}''|$,
\begin{equation}
\mathcal{V}(a,b,\mathcal{V}',\mathcal{V}'')=\big\{(\bar{\mathcal{V}}',\bar{\mathcal{V}}'')\big||\bar{\mathcal{V}}'|=a,|\bar{\mathcal{V}}''|=b,(\bar{\mathcal{V}}',\bar{\mathcal{V}}'')\subseteq
(\mathcal{V}',\mathcal{V}'') \big\}
\end{equation}
and for  $\mathbf{H}\in\mathbb{F}_2^{|\mathcal{V}''|\times
|\mathcal{V}'|}$,
\begin{eqnarray}
\mathcal{V}\left(\mathbf{H},\mathcal{V}',\mathcal{V}''\right)\!\!\!\!\!\!\!\!&&=\big\{(\bar{\mathcal{V}}',\bar{\mathcal{V}}'')\big|\rank(\mathbf{H}_{\bar{\mathcal{V}}',\bar{\mathcal{V}}''}[1])=|\bar{\mathcal{V}}'|=|\bar{\mathcal{V}}''|=\rank(\mathbf{H})\nonumber\\
&&{~~~~~}\mbox{ where
}\mathbf{H}_{\mathcal{V}',\mathcal{V}''}[1]=\mathbf{H},(\bar{\mathcal{V}}',\bar{\mathcal{V}}'')\subseteq
(\mathcal{V}',\mathcal{V}'')\big\},
\end{eqnarray}
where
$\mathcal{V}\left(\mathbf{H},\mathcal{V}',\mathcal{V}''\right)=\phi$
if $\rank(\mathbf{H})=0$. The set
$\mathcal{V}(a,b,\mathcal{V}',\mathcal{V}'')$ consists of all
$(\bar{\mathcal{V}}',\bar{\mathcal{V}}'')\subseteq(\mathcal{V}',\mathcal{V}'')$ such that the number of
nodes in $\bar{\mathcal{V}}'$ and the number of nodes in
$\bar{\mathcal{V}}''$ are equal to $a$ and $b$, respectively. The
set $\mathcal{V}\left(\mathbf{H},\mathcal{V}',\mathcal{V}''\right)$
consists of all $(\bar{\mathcal{V}}',\bar{\mathcal{V}}'')\subseteq(\mathcal{V}',\mathcal{V}'')$ such that
$\mathbf{H}_{\bar{\mathcal{V}}',\bar{\mathcal{V}}''}[1]$ is a
full-rank matrix and has the same rank as $\mathbf{H}$, where
$\mathbf{H}_{\mathcal{V}',\mathcal{V}''}[1]=\mathbf{H}$.

\section{Upper Bound} \label{sec:converse}
In this section, we derive a general cut-set upper bound, which will
be used to show the converses in Section \ref{sec:achievability}.
\subsection{Cut-set Upper Bound}
We show that any sequence of $(2^{nR_1},\cdots,2^{nR_K};n)$ codes
with $P^{(n)}_{e,k}\to 0$ for all $k\in\{1,\cdots,K\}$ satisfies the
rate constraints in the following theorem.

\begin{theorem} \label{THM:cut_set}
Suppose a linear binary field relay network. For a cut $\Omega$, the
set of achievable rates $(R_1,\cdots,R_K)$ is upper bounded by
\begin{equation}
\sum_{k\in\mathcal{K}_{\Omega}}R_k\leq
\mathbb{E}(\rank(\mathbf{H}_{\Omega}[1])).
\label{EQ:general_cut_set}
\end{equation}
\end{theorem}
\begin{proof}
Let us define
$\mathcal{W}_{\mathcal{K}_{\Omega}}=\{W_k\big|k\in\mathcal{K}_{\Omega}\}$.
We further define a length-$n$ sequence $a^n$ to denote
$\{a[1],\cdots,a[n]\}$. Then
\begin{eqnarray}
n\sum_{k\in\mathcal{K}_{\Omega}}R_k\!\!\!\!\!\!\!&&=H(\mathcal{W}_{\mathcal{K}_{\Omega}})\nonumber\\
&&=I(\mathcal{W}_{\mathcal{K}_{\Omega}};\mathcal{Y}^n_{\mathcal{D}_{\Omega}},\mathbf{H}_1^n,\cdots,\mathbf{H}_M^n)+H(\mathcal{W}_{\mathcal{K}_{\Omega}}|\mathcal{Y}^n_{\mathcal{D}_{\Omega}},\mathbf{H}_1^n,\cdots,\mathbf{H}_M^n)\nonumber\\
&&\overset{(a)}{\leq}I(\mathcal{W}_{\mathcal{K}_{\Omega}};\mathcal{Y}^n_{\mathcal{D}_{\Omega}},\mathbf{H}_1^n,\cdots,\mathbf{H}_M^n)+n\epsilon_n\nonumber\\
&&\overset{(b)}{=}I(\mathcal{W}_{\mathcal{K}_{\Omega}};\mathcal{Y}^n_{\mathcal{D}_{\Omega}}|\mathbf{H}_1^n,\cdots,\mathbf{H}_M^n)+n\epsilon_n\nonumber\\
&&\overset{(c)}{\leq}I(\mathcal{W}_{\mathcal{K}_{\Omega}};\mathcal{Y}^n_{\Omega_D}|\mathbf{H}_1^n,\cdots,\mathbf{H}_M^n)+n\epsilon_n\nonumber\\
&&\overset{(d)}{\leq}H(\mathcal{W}_{\mathcal{K}_{\Omega}}|\mathcal{X}^n_{\Omega\setminus \Omega'},\mathbf{H}_1^n,\cdots,\mathbf{H}_M^n)-H(\mathcal{W}_{\mathcal{K}_{\Omega}}|\mathcal{X}^n_{\Omega\setminus\Omega'},\mathcal{Y}^n_{\Omega_D},\mathbf{H}_1^n,\cdots,\mathbf{H}_M^n)+n\epsilon_n\nonumber\\
&&=I(\mathcal{W}_{\mathcal{K}_{\Omega}};\mathcal{Y}^n_{\Omega_D}|\mathcal{X}^n_{\Omega\setminus\Omega'},\mathbf{H}_1^n,\cdots,\mathbf{H}_M^n)+n\epsilon_n\nonumber\\
&&\leq H(\mathcal{Y}^n_{\Omega_D}|\mathcal{X}^n_{\Omega\setminus\Omega'},\mathbf{H}_1^n,\cdots,\mathbf{H}_M^n)+n\epsilon_n\nonumber\\
&&\overset{(e)}{=}\sum_{t=1}^nH(\mathcal{Y}_{\Omega_D}[t]|\mathcal{X}_{\Omega\setminus\Omega'}[t],\mathbf{H}_1[t],\cdots,\mathbf{H}_M[t])+n\epsilon_n\nonumber\\
&&\overset{(f)}{\leq}n\mathbb{E}(\rank(\mathbf{H}_{\Omega}[1]))+n\epsilon_n,
\label{EQ:general_converse_proof}
\end{eqnarray}
where $\epsilon_n> 0$ satisfies $\epsilon_n\to 0$ as $n\to\infty$. Notice that $(a)$ holds from Fano's inequality, $(b)$ holds since the messages are independent of
channels, $(c)$ holds since
$\mathcal{W}_{\mathcal{K}_{\Omega}}-\left(\mathcal{Y}^n_{\Omega_D},\mathbf{H}_1^n,\cdots,\mathbf{H}_M^n\right)-\mathcal{Y}^n_{\mathcal{D}_{\Omega}}$
forms a Markov chain, $(d)$ holds since
$\mathcal{W}_{\mathcal{K}_{\Omega}}$ is independent of
$\mathcal{X}^n_{\Omega\setminus\Omega'},\mathbf{H}_1^n,\cdots,\mathbf{H}_M^n$ and conditioning reduces entropy, $(e)$ holds since channels
are memoryless, and $(f)$ holds with equality if
$\mathcal{X}_{\Omega_S}[t]$ is uniformly distributed over
$\mathbb{F}^{|\Omega_S|}_2$. Therefore, we have
(\ref{EQ:general_cut_set}), which completes the proof.
\end{proof}

\begin{figure}[t!]
  \begin{center}
  \scalebox{0.9}{\includegraphics{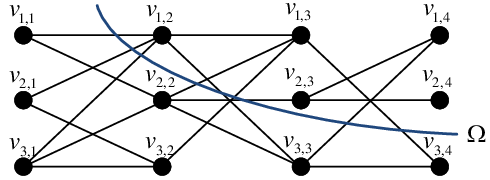}}
  \caption{Example of the cut-set upper bound, where the solid lines mean that the corresponding channels become ones with non-zero probabilities.}
  \label{FIG:example_cut_set}
  \end{center}
\end{figure}

Theorem \ref{THM:cut_set} shows that the aggregate rate of the S--D
pairs divided by a cut is upper bounded by the average rank of the
channel matrix constructed by the cut.
\begin{example}[Cut-set Upper Bound]
Consider the cut
$\Omega=\{v_{1,1},v_{2,1},v_{3,1},v_{2,2},v_{3,2},v_{3,3},v_{3,4}\}$
in Fig. \ref{FIG:example_cut_set}. Then we obtain
$\mathcal{D}_{\Omega}=\{v_{1,4},v_{2,4}\}$,
$\mathcal{S}_{\Omega}=\{v_{1,1},v_{2,1}\}$,
$\Omega_{D}=\{v_{2,3},v_{1,4}\}$, $\Omega_{S}=\{v_{2,2},v_{3,3}\}$,
and $\mathbf{H}_{\Omega}[1]$ is given by
$[[h_{2,2,2}[1],0]^T,[0,h_{1,3,3}[1]]^T]^T$. Therefore, $R_1+R_2$ is
upper bounded by
$\mathbb{E}(\operatorname{rank}(\mathbf{H}_{\Omega}[1]))=p_{2,2,2}+p_{1,3,3}$.
\end{example}


\subsection{Rate Bounds for Single-hop and Multi-hop Networks}
In this subsection, we obtain useful rate upper bounds from Theorem
\ref{THM:cut_set}, which will be used to show the converses in
Corollaries \ref{CO:capacity_single} and \ref{CO:capacity_multi}.
Let us first consider single-hop networks, that is $M=1$. If we set
$\Omega=\{v_{k,1}\}$, then $\sum_{i\in\mathcal{K}_{\Omega}}R_i=R_k$
and $\mathbf{H}_{\Omega}[t]=h_{k,k,1}[t]$. Thus, we obtain
\begin{equation}
R_k\leq p_{k,k,1} \label{EQ:converse_single}
\end{equation}
for all $k\in\{1,\cdots,K\}$.
Let us now consider multi-hop networks, that is $M\geq2$. By setting
$\Omega=\cup_{i\in\{1,\cdots,m\}}\mathcal{V}_i$, we have
$\sum_{k\in\mathcal{K}_{\Omega}}R_k=R_{\operatorname{sum}}$ and
$\mathbf{H}_{\Omega}[t]=\mathbf{H}_m[t]$, where
$m\in\{1,\cdots,M\}$. Hence, we obtain
\begin{equation}
R_{\operatorname{sum}}\leq\min_{m\in\{1,\cdots,M\}}\mathbb{E}(\rank(\mathbf{H}_m[1]))
\label{EQ:converse_multi}
\end{equation}
or equivalently
$R_{\operatorname{sum}}\leq\mathbb{E}(\rank(\mathbf{H}_{m_0}[1]))$.
\section{Achievability} \label{sec:achievability}
In this section, we propose transmission schemes and derive their
achievable rate regions.

\subsection{Achievability for $M=1$} \label{subsec:achievability_single}
Consider a single-hop network, that is $M=1$. As mentioned in Introduction, each source can transmit one bit without interference
by using two particular instances $\mathbf{H}_1^{(1)}$ and
$\mathbf{H}_1^{(2)}$ jointly such that
$\mathbf{H}_1^{(1)}+\mathbf{H}_1^{(2)}=\mathbf{I}$.
The proposed encoding makes such pairing possible.
\subsubsection{Proposed scheme}
Let us divide a block into two sub-blocks having length $n/2$ for
each sub-block. For $\mathbf{H}_1\in \mathbb{F}^{K\times K}_2$,
define $\mathcal{T}_b(\mathbf{H}_1)$ as the set of time indices
of the $b$-th sub-block whose channel instances are equal to
$\mathbf{H}_1$, where $b\in\{1,2\}$. We further define
\begin{equation}
n(\mathbf{H}_1)=c_1^{-1}nR\min\{\Pr(\mathbf{H}_1[1]=\mathbf{H}_1),\Pr(\mathbf{H}_1[1]=\mathbf{H}_1+\mathbf{I})\},
\label{EQ:n_H}
\end{equation}
where
\begin{equation}
c_1=\sum_{\mathbf{H}_1\in\mathbb{F}_2^{K\times
K}}\min\{\mathbf{H}_1[1]=\Pr(\mathbf{H}_1),\Pr(\mathbf{H}_1[1]=\mathbf{H}_1+\mathbf{I})\}.
\end{equation}

The detailed encoding is as follows.
\begin{itemize}
\item (Encoding of the first sub-block)

For all $\mathbf{H}_1\in \mathbb{F}_2^{K\times K}$, declare an error
if $|\mathcal{T}_1(\mathbf{H}_1)|<n(\mathbf{H}_1)$, otherwise
each source transmits $n(\mathbf{H}_1)$ information bits using the
time indices in $\mathcal{T}_1(\mathbf{H}_1)$.
\item (Encoding of the second sub-block)

For all $\mathbf{H}_1\in \mathbb{F}_2^{K\times K}$, declare an error
if $|\mathcal{T}_2(\mathbf{H}_1)|<n(\mathbf{H}_1)$, otherwise
each source retransmits $n(\mathbf{H}_1)$ information bits that were
transmitted during $\mathcal{T}_1(\mathbf{H}_1+\mathbf{I})$
using the time indices in $\mathcal{T}_2(\mathbf{H}_1)$.
\end{itemize}

Notice that, since each source transmits
$\sum_{\mathbf{H}_1\in\mathbb{F}^{K\times K}_2}n(\mathbf{H}_1)$
information bits during $n$ channel uses, the transmission rates are
given by
$R_1=\cdots=R_K=\frac{1}{n}\sum_{\mathbf{H}_1\in\mathbb{F}^{K\times
K}_2}n(\mathbf{H}_1)=R$.
Let $s_k(i)$ denote the $i$-th information bit of the $k$-th source,
where $i=\{1,\cdots,nR\}$. Let $t_1(i)$ and $t_2(i)$ denote the time
indices over which $s_k(i)$ was transmitted. Then the detailed
decoding is as follows.
\begin{itemize}
\item (Decoding)

For $i\in\{1,\cdots,nR\}$, the $k$-th destination sets
$\hat{s}_k(i)=y_{k,1}[t_1(i)]+y_{k,1}[t_2(i)]$.
\end{itemize}


\subsubsection{Achievable rate region}
We derive the achievable rate region of the proposed scheme. Let
$E_b$ denote the event such that
$|\mathcal{T}_b(\mathbf{H}_1)|<n(\mathbf{H}_1)$ for any
$\mathbf{H}_1\in\mathbb{F}^{K\times K}$, where $b\in\{1,2\}$. The
following lemma shows that there is no error if $(E_1\cup
E_2)^c$ occurs.

\begin{lemma} \label{LEM:P_e_single}
Suppose a linear binary field relay network with $M=1$. The
probability of error is upper bounded by
\begin{equation}
P^{(n)}_{e,k}\leq \Pr(E_1)+\Pr(E_2) \label{EQ:p_n_e_k_upper}
\end{equation}
for all $k\in\{1,\cdots,K\}$.
\end{lemma}
\begin{proof}
The proof is in Appendix I.
\end{proof}

Then the remaining thing is to derive $R$ that guarantees
$P^{(n)}_{e,k}\to 0$ as $n\to\infty$. The following theorem
characterizes such $R$.

\begin{theorem} \label{THM:achievable_rate_single}
Suppose a linear binary field relay network with $M=1$. Then
\begin{equation}
R_k=\frac{1}{2}\sum_{\mathbf{H}_1\in\mathbb{F}_2^{K\times
K}}\min\{\Pr(\mathbf{H}_1[1]=\mathbf{H}_1),\Pr(\mathbf{H}_1[1]=\mathbf{H}_1+\mathbf{I})\}
\label{EQ:rate_single}
\end{equation}
is achievable for all  $k\in\{1,\cdots,K\}$.
\end{theorem}
\begin{proof}
Let us consider $|\mathcal{T}_b(\mathbf{H}_1)|$. By the weak law
of large numbers \cite{Csiszar:81}, there exists a sequence
$\epsilon_n\to0$ as $n\to\infty$ such that the probability
\begin{equation}
|\mathcal{T}_b(\mathbf{H}_1)|\geq\frac{n}{2}(\Pr(\mathbf{H}_1[1]=\mathbf{H}_1)-\delta_n)\mbox{
for all }\mathbf{H}_1
\end{equation}
is greater than or equal to $1-\epsilon_n$, where $\delta_n\to 0$ as
$n\to\infty$. This indicates that $\Pr(E_b)\leq \epsilon_n$ if
$n(\mathbf{H}_1)\leq\frac{n}{2}(\Pr(\mathbf{H}_1[1]=\mathbf{H}_1)-\delta_n)$
for all $\mathbf{H}_1$. Hence, from (\ref{EQ:n_H}), if
\begin{equation}
R\leq\frac{c_1(\Pr(\mathbf{H}_1[1]=\mathbf{H}_1)-\delta_n)}{2\min\{\Pr(\mathbf{H}_1[1]=\mathbf{H}_1),\Pr(\mathbf{H}_1[1]=\mathbf{H}_1+\mathbf{I})\}}
\end{equation}
for all $\mathbf{H}_1$, then $P^{(n)}_{e,k}\leq 2\epsilon_n$, where
we use the result of Lemma \ref{LEM:P_e_single}. Thus we set
$R=\frac{c_1}{2}(1-\delta^*_n)$, where
$\delta^*_n=\frac{\delta_n}{\min_{\mathbf{H}'_1\in\mathbb{F}_2^{K\times
K}}\{\Pr(\mathbf{H}_1[1]=\mathbf{H}'_1)\}}$, which converges to zero
as $n\to\infty$. In conclusion, (\ref{EQ:rate_single}) is achievable
for all $k\in\{1,\cdots,K\}$, which completes the proof.
\end{proof}

\begin{corollary} \label{CO:capacity_single}
Suppose a linear binary field relay network with $M=1$. If
$p_{k,k,1}= 1/2$ for all $k\in\{1,\cdots,K\}$, the capacity region
is given by all rate tuples $(R_1,\cdots,R_K)$ satisfying
\begin{equation}
R_k\leq\frac{1}{2}
\end{equation}
for all $k\in\{1,\cdots,K\}$.
\end{corollary}
\begin{proof}
Note that
$\Pr(\mathbf{H}_1[1]=\mathbf{H}_1)=\Pr(\mathbf{H}_1[1]=\mathbf{H}_1+\mathbf{I})$
for all $\mathbf{H}_1$ if $p_{k,k,1}= 1/2$. Hence, from
(\ref{EQ:rate_single}),
$R_k=\frac{1}{2}\sum_{\mathbf{H}_1\in\mathbb{F}_2^{K\times
K}}\Pr(\mathbf{H}_1[1]=\mathbf{H}_1)=\frac{1}{2}$ is achievable for
all $k\in\{1,\cdots,K\}$. Note that the achievable rate region
coincides with the upper bound in (\ref{EQ:converse_single}), which
provides the capacity region. Therefore, Corollary
\ref{CO:capacity_single} holds.
\end{proof}

\begin{remark} \label{RE:q_ary_single}
Corollary \ref{CO:capacity_single} can be directly extended to a
general linear finite field relay network in which inputs, outputs,
and channels are in $\mathbb{F}_q$ and channels are i.i.d. uniformly
distributed over $\mathbb{F}_q$. Specifically, the capacity region
is given by all rate tuples $(R_1,\cdots,R_K)$ satisfying $R_k\leq
\frac{1}{2}\log q$ for all $k\in\{1,\cdots,K\}$.
\end{remark}

Corollary \ref{CO:capacity_single} shows that all S--D pairs can simultaneously achieve the capacity of the
\emph{point-to-point channel assuming no interference} if the
direct channels are uniformly distributed. This result also shows
that the \emph{max-flow min-cut theorem} holds for a certain class
of channel distributions. Similar to the Gaussian interference
channel in which $1/2$ degrees of freedom is achievable for each
S--D pair \cite{Viveck1:08}, each source can transmit data to its
destination with a non-vanishing rate even as $K$ tends to infinity.

\begin{example} [$2$--$2$ network]
Consider the case where $K=2$ and $M=1$ with $p_{j,i,1}=1/2$ for all
$i$ and $j$. If we use each channel instance separately, then
$R_{\operatorname{sum}}\leq 13/16$ is achievable. However, the
proposed scheme achieves $R_{\operatorname{sum}}\leq1$. More
specifically, $R_1\leq1/2$ and $R_2\leq1/2$ are achievable, which is
the capacity region of this network.
\end{example}

\subsection{Achievability for $M\geq 2$} \label{subsec:achievability_multi}
Consider a multi-hop network, that is $M\geq 2$. As mentioned in
Introduction, each source can transmit one bit to its destination
without interference through particular instances from
$\mathbf{H}_1$ to $\mathbf{H}_M$ such that
\begin{equation}
\mathbf{H}_M\mathbf{H}_{M-1}\cdots\mathbf{H}_1=\mathbf{I}.
\label{EQ:condition_multi}
\end{equation}
Due to network topologies and channel distributions, however, some
instances will be rank-deficient and it is impossible to find a
series of pairs satisfying (\ref{EQ:condition_multi}) by using
rank-deficient instances. Furthermore, a series of pairs satisfying
(\ref{EQ:condition_multi}) is not unique and the number of possible
pairing increases exponentially as the number of nodes in a layer or
the number of layers increases. Hence, we first reduce the size of
effective channels by transmitting and receiving using subsets of
nodes at each hop such that the average ranks are balanced between
hops and their instances have full-rank. Then we randomize a series
of pairs based on these effective channels.

\subsubsection{Construction of effective channels}
Recall that the $m_0$-th hop becomes a bottleneck for the entire
multi-hop transmission, which can be verified from
(\ref{EQ:converse_multi}). Hence, we select
$\mathcal{V}_{m,\operatorname{tx}}[t]\subseteq\mathcal{V}_m$ and
$\mathcal{V}_{m,\operatorname{rx}}[t]\subseteq\mathcal{V}_{m+1}$
randomly such that
\begin{equation}
(\mathcal{V}_{m,\operatorname{tx}}[t],\mathcal{V}_{m,\operatorname{rx}}[t])\in\mathcal{V}(K_{m_0},K_{m_0+1},\mathcal{V}_m,\mathcal{V}_{m+1})
\end{equation}
with equal probabilities (or in
$\mathcal{V}(K_{m_0+1},K_{m_0},\mathcal{V}_m,\mathcal{V}_{m+1})$).
Notice that this is possible since the considered network has a
minimum-dimensional bottleneck-hop. Because the maximum number of
bits transmitted at the $m$-th hop is limited by
$\rank(\mathbf{H}_{\mathcal{V}_{m,\operatorname{tx}}[t],\mathcal{V}_{m,\operatorname{rx}}[t]}[t])$,
we further select
$\bar{\mathcal{V}}_{m,\operatorname{tx}}[t]\subseteq\mathcal{V}_{m,\operatorname{tx}}[t]$
and
$\bar{\mathcal{V}}_{m,\operatorname{rx}}[t]\subseteq\mathcal{V}_{m,\operatorname{rx}}[t]$
randomly such that
\begin{equation}
(\bar{\mathcal{V}}_{m,\operatorname{tx}}[t],\bar{\mathcal{V}}_{m,\operatorname{rx}}[t])\in\mathcal{V}(\mathbf{H}_{\mathcal{V}_{m,\operatorname{tx}}[t],\mathcal{V}_{m,\operatorname{rx}}[t]}[t],\mathcal{V}_{m,\operatorname{tx}}[t],\mathcal{V}_{m,\operatorname{rx}}[t])
\end{equation}
with equal probabilities. For each time $t$, the nodes in
$\bar{\mathcal{V}}_{m,\operatorname{tx}}[t]$ transmit and the nodes
in $\bar{\mathcal{V}}_{m,\operatorname{rx}}[t]$ receive through
their effective channel
$\mathbf{H}_{\bar{\mathcal{V}}_{m,\operatorname{tx}}[t],\bar{\mathcal{V}}_{m,\operatorname{rx}}[t]}[t]$
at the $m$-th hop. Then information bits can be transmitted using
particular time indices $t_1,\cdots,t_M$ such that
$\bar{\mathcal{V}}_{1,\operatorname{tx}}[t_1]=\bar{\mathcal{V}}_{M,\operatorname{rx}}[t_M]$,
$\bar{\mathcal{V}}_{m,\operatorname{tx}}[t_{m}]=\bar{\mathcal{V}}_{m-1,\operatorname{rx}}[t_{m-1}]$
for all $m\in\{2,\cdots,M\}$, and
\begin{equation}
\mathbf{H}_{\bar{\mathcal{V}}_{M,\operatorname{tx}}[t_M],\bar{\mathcal{V}}_{M,\operatorname{rx}}[t_M]}[t_M]\cdots\mathbf{H}_{\bar{\mathcal{V}}_{1,\operatorname{tx}}[t_1],\bar{\mathcal{V}}_{1,\operatorname{rx}}[t_1]}[t_1]=\mathbf{I},
\end{equation}
which guarantees interference-free reception at the destinations.
It is possible to construct those pairs because effective channels
are always invertible\footnote{We do not use the effective channels having all
zeros, which give zero rate.}.
Let $\mathcal{F}_i$ be the set of all full-rank matrices in $\mathbb{F}_2^{i\times i}$, where $i\in\{1,\cdots,K_{\min}\}$.
The following lemma shows useful probability distributions, which will be used to derive the achievable
rate region of the proposed scheme.

\begin{lemma} \label{LEM:mapping1}
Suppose a linear binary field relay network with $M\geq 2$. If the
network has a minimum-dimensional bottleneck-hop and $p_{j,i,m}=p$
for all $i$, $j$, and $m$, then the following probabilities hold:
\begin{enumerate}
\item For $\mathbf{H}\in \mathbb{F}_2^{K_{m_0}\times K_{m_0+1}}$ (or $\mathbb{F}_2^{K_{m_0+1}\times K_{m_0}}$),
\begin{eqnarray}
\Pr(\mathbf{H}_{\mathcal{V}_{m,\operatorname{tx}}[t],\mathcal{V}_{m,\operatorname{rx}}[t]}[t]=\mathbf{H})=p^{u}(1-p)^{K_{m_0+1}K_{m_0}-u},
\label{Eq:p_H}
\end{eqnarray}
where $u$ is the number of ones in $\mathbf{H}$.
\item For $\mathbf{G}\in \mathcal{F}_i$,
\begin{eqnarray}
&&\Pr(\mathbf{H}_{\bar{\mathcal{V}}_{m,\operatorname{tx}}[t],\bar{\mathcal{V}}_{m,\operatorname{rx}}[t]}[t]=\mathbf{G})\nonumber\\
&&=\sum_{\underset{\mathcal{V}(i,i,\mathcal{V}_{m_0},\mathcal{V}_{m_0+1})}{(\mathcal{V}',\mathcal{V}'')\in}}\sum_{\mathbf{H}\in\mathcal{H}^F_{\mathcal{V}_{m_0},\mathcal{V}_{m_0+1}}(\mathbf{G},\mathcal{V}',\mathcal{V}'')}\frac{\Pr(\mathbf{H}_{\mathcal{V}_{m,\operatorname{tx}}[t],\mathcal{V}_{m,\operatorname{rx}}[t]}[t]=\mathbf{H})}{|\mathcal{V}(\mathbf{H},\mathcal{V}_{m_0},\mathcal{V}_{m_0+1})|},
\label{EQ:p_G}
\end{eqnarray}
where
$\Pr(\mathbf{H}_{\mathcal{V}_{m,\operatorname{tx}}[t],\mathcal{V}_{m,\operatorname{rx}}[t]}[t]=\mathbf{H})$
is given by (\ref{Eq:p_H}). If $p=1/2$, we have
\begin{equation}
\Pr(\mathbf{H}_{\bar{\mathcal{V}}_{m,\operatorname{tx}}[t],\bar{\mathcal{V}}_{m,\operatorname{rx}}[t]}[t]=\mathbf{G})=2^{-K_{m_0+1}K_{m_0}}\frac{N_{K_{m_0+1},
K_{m_0}}(i)}{N_{i,i}(i)},
\label{EQ:p_G_equal_prob}
\end{equation}
where $N_{a,b}(c)$ is the number of channel matrices in
$\mathbb{F}_2^{a\times b}$ having rank $c$.

\item For $\mathbf{G}\in \mathcal{F}_i$ and $(\mathcal{V}'_m,\mathcal{V}'_{m+1})\in \mathcal{V}(i,i,\mathcal{V}_m,\mathcal{V}_{m+1})$,
\begin{eqnarray}
&&\Pr(\mathbf{H}_{\bar{\mathcal{V}}_{m,\operatorname{tx}}[t],\bar{\mathcal{V}}_{m,\operatorname{rx}}[t]}[t]=\mathbf{G},\bar{\mathcal{V}}_{m,\operatorname{tx}}[t]=\mathcal{V}_m',\bar{\mathcal{V}}_{m,\operatorname{rx}}[t]=\mathcal{V}_{m+1}')\nonumber\\
&&=\frac{\Pr(\mathbf{H}_{\bar{\mathcal{V}}_{m,\operatorname{tx}}[t],\bar{\mathcal{V}}_{m,\operatorname{rx}}[t]}[t]=\mathbf{G})}{{K_m\choose
i}{K_{m+1}\choose i}},
\label{EQ:p_3}
\end{eqnarray}
where
$\Pr(\mathbf{H}_{\bar{\mathcal{V}}_{m,\operatorname{tx}}[t],\bar{\mathcal{V}}_{m,\operatorname{rx}}[t]}[t]=\mathbf{G})$
is given by (\ref{EQ:p_G}).
\end{enumerate}
\end{lemma}
\begin{proof}
The proof is in Appendix II.
\end{proof}

Note that the probabilities in (\ref{Eq:p_H}) to (\ref{EQ:p_3}) are
the same for all $m$ and $t$. For notational simplicity, we use the shorthand notation $P_G(\mathbf{G})$ to denote $\Pr(\mathbf{H}_{\bar{\mathcal{V}}_{m,\operatorname{tx}}[t],\bar{\mathcal{V}}_{m,\operatorname{rx}}[t]}[t]=\mathbf{G})$. That is, for $\mathbf{G}\in\mathcal{F}_i$,
\begin{eqnarray}
P_G(\mathbf{G})=\sum_{\underset{\mathcal{V}(i,i,\mathcal{V}_{m_0},\mathcal{V}_{m_0+1})}{(\mathcal{V}',\mathcal{V}'')\in}}\sum_{\mathbf{H}\in\mathcal{H}^F_{\mathcal{V}_{m_0},\mathcal{V}_{m_0+1}}(\mathbf{G},\mathcal{V}',\mathcal{V}'')}\frac{P_H(\mathbf{H})}{|\mathcal{V}(\mathbf{H},\mathcal{V}_{m_0},\mathcal{V}_{m_0+1})|},
\label{eq:p_2_G}
\end{eqnarray}
where $P_H(\mathbf{H})=p^{u}(1-p)^{K_{m_0+1}K_{m_0}-u}$ and $u$ is the number of ones in $\mathbf{H}$.

\subsubsection{Proposed scheme}
Divide a block into $B+M-1$ sub-blocks having length $n_B$ for each
sub-block, where $n_B=\frac{n}{B+M-1}$. Since block encoding
and relaying are applied over $M$ hops, the number of effective
sub-blocks is equal to $B$. Thus, the overall rate is given by
$\frac{B}{B+M-1}R_k$. As $n\to\infty$, the fractional rate loss
$1-\frac{B}{B+M-1}$ will be negligible because we can make both
$n_B$ and $B$ large enough. For simplicity, we omit the sub-block
index in describing the proposed scheme.

We divide $M$ hops into two parts, the first $N$ hops and the rest of the $M-N$ hops, where $N\in\{1,\cdots,M-1\}$.
Then, for $\mathbf{G}\in\mathcal{F}_i$, define
\begin{eqnarray}
P_{\alpha}(\mathbf{G})\!\!\!\!\!\!\!\!&&=\zeta_i^{-(N-1)}\sum_{\underset{\mathbf{G}_{N}\cdots\mathbf{G}_1=\mathbf{G}}{\mathbf{G}_1,\cdots,\mathbf{G}_N\in \mathcal{F}_i,}}\prod_{m=1}^{N}P_G(\mathbf{G}_m),\label{EQ:p_A}\\
P_{\beta}(\mathbf{G})\!\!\!\!\!\!\!\!&&=\zeta_i^{-(M-N-1)}\sum_{\underset{\mathbf{G}_{M}\cdots\mathbf{G}_{N+1}=\mathbf{G}}{\mathbf{G}_{N+1},\cdots,\mathbf{G}_M\in \mathcal{F}_i,}}\prod_{m=N+1}^{M}P_G(\mathbf{G}_m),
\label{EQ:p_B}
\end{eqnarray}
and
\begin{equation}
n(\mathbf{G})=c_2^{-1} n_B R\min\{P_{\alpha}(\mathbf{G}),P_{\beta}(\mathbf{G}^{-1})\},
\end{equation}
where $\zeta_i=\sum_{\mathbf{G}'\in\mathcal{F}_i}P_G(\mathbf{G}')$ and $c_2=\frac{1}{K}\sum_{j=1}^{K_{\min}}j\sum_{\mathbf{G}'\in\mathcal{F}_j}\min\{P_{\alpha}(\mathbf{G}'),P_{\beta}(\mathbf{G}'^{-1})\}$.
We further define
\begin{eqnarray}
n_{\alpha}(\mathbf{G}_1,\cdots,\mathbf{G}_N)\!\!\!\!\!\!\!\!&&=c_2^{-1} n_B R \zeta_i^{-(N-1)}\prod_{m=1}^{N}\left(P_G(\mathbf{G}_m)-\Delta_{\alpha}(\mathbf{G}_1,\cdots,\mathbf{G}_N)\right),\label{eq:n_a1}\\
n_{\beta}(\mathbf{G}_{N+1},\cdots,\mathbf{G}_M)\!\!\!\!\!\!\!\!&&=c_2^{-1} n_B R \zeta_i^{-(M-N-1)}\prod_{m=N+1}^{M}\left(P_G(\mathbf{G}_m)-\Delta_{\beta}(\mathbf{G}_{N+1},\cdots,\mathbf{G}_M)\right)\label{eq:n_a2},
\end{eqnarray}
where $\mathbf{G}_1,\cdots,\mathbf{G}_M\in \mathcal{F}_i$.
Here, $\Delta_{\alpha}(\mathbf{G}_1,\cdots,\mathbf{G}_N)\geq0$ and $\Delta_{\beta}(\mathbf{G}_{N+1},\cdots,\mathbf{G}_M)\geq0$ are set such that
\begin{equation}
\sum_{\underset{\mathbf{G}'_{N}\cdots\mathbf{G}'_1=\mathbf{G}}{\mathbf{G}'_1,\cdots,\mathbf{G}'_N\in \mathcal{F}_i,}}n_{\alpha}(\mathbf{G}'_1,\cdots,\mathbf{G}'_N)=\sum_{\underset{\mathbf{G}'_{M}\cdots\mathbf{G}'_{N+1}=\mathbf{G}^{-1}}{\mathbf{G}'_{N+1},\cdots,\mathbf{G}'_M\in \mathcal{F}_i,}}n_{\beta}(\mathbf{G}'_{N+1},\cdots,\mathbf{G}'_M)=n(\mathbf{G})
\end{equation}
is satisfied for all $\mathbf{G}\in \mathcal{F}_i$.

For a given $\mathbf{G}\in\mathcal{F}_i$, the proposed scheme transmits $i\times n_{\alpha}(\mathbf{G}_1,\cdots,\mathbf{G}_N)$ bits through a series of effective channels $\mathbf{G}_1$ to $\mathbf{G}_N$ satisfying $\mathbf{G}_N\cdots\mathbf{G}_1=\mathbf{G}$ for all $\mathbf{G}_1,\cdots,\mathbf{G}_N\in\mathcal{F}_i$.
Hence a total of $i\sum_{\underset{\mathbf{G}_{N}\cdots\mathbf{G}_1=\mathbf{G}}{\mathbf{G}_1,\cdots,\mathbf{G}_N\in \mathcal{F}_i,}}n_{\alpha}(\mathbf{G}_1,\cdots,\mathbf{G}_N)=i\times n(\mathbf{G})$ bits are transmitted.
Then these $i\times n(\mathbf{G})$ received bits are transmitted through $\mathbf{G}_{N+1}$ to $\mathbf{G}_M$ satisfying $\mathbf{G}_M\cdots\mathbf{G}_{N+1}=\mathbf{G}^{-1}$ for all $\mathbf{G}_{N+1},\cdots,\mathbf{G}_M\in\mathcal{F}_i$.
More specifically, $i\times n_{\beta}(\mathbf{G}_{N+1},\cdots,\mathbf{G}_M)$ bits are transmitted through $\mathbf{G}_{N+1}$ to $\mathbf{G}_M$ and, as a result, a total of $i\sum_{\underset{\mathbf{G}_{M}\cdots\mathbf{G}_{N+1}=\mathbf{G}^{-1}}{\mathbf{G}_{N+1},\cdots,\mathbf{G}_M\in \mathcal{F}_i,}}n_{\beta}(\mathbf{G}_{N+1},\cdots,\mathbf{G}_M)=i\times n(\mathbf{G})$ bits are transmitted.
Let
\begin{equation}
n_m(\mathbf{G}_m)=\begin{cases}\sum_{\mathbf{G}_1,\cdots,\mathbf{G}_{m-1},\mathbf{G}_{m+1},\cdots,\mathbf{G}_N\in\mathcal{F}_i}n_{\alpha}(\mathbf{G}_1,\cdots,\mathbf{G}_N)&\mbox { for }m\in\{1,\cdots,N\},\\
\sum_{\mathbf{G}_{N+1},\cdots,\mathbf{G}_{m-1},\mathbf{G}_{m+1},\cdots,\mathbf{G}_M\in\mathcal{F}_i}n_{\beta}(\mathbf{G}_{N+1},\cdots,\mathbf{G}_M)&\mbox { for }m\in\{N+1,\cdots,M\},
\end{cases}
\label{EQ:n_m}
\end{equation}
where $\mathbf{G}_m\in\mathcal{F}_i$.
Then $i\times n_m(\mathbf{G}_m)$ is the total number of bits that are transmitted through $\mathbf{G}_m$ at the $m$-th hop.
Define
$\mathcal{T}_m(\mathbf{G}_m, \mathcal{V}'_m,\mathcal{V}'_{m+1})$ as
the set of time indices of the sub-block at the $m$-th hop
satisfying
$\bar{\mathcal{V}}_{\operatorname{tx},m}[t]=\mathcal{V}'_m$,
$\bar{\mathcal{V}}_{\operatorname{rx},m}[t]=\mathcal{V}'_{m+1}$, and
$\mathbf{H}_{\mathcal{V}'_m,\mathcal{V}'_{m+1}}[t]=\mathbf{G}_m$, where $\mathbf{G}_m\in \mathcal{F}_i$
and $(\mathcal{V}'_m,\mathcal{V}'_{m+1})\in
\mathcal{V}(i,i,\mathcal{V}_m,\mathcal{V}_{m+1})$.
For all $i\in \{1,\cdots,K_{\min}\}$, the detailed encoding and relaying are as follows.

\begin{itemize}
\item (Encoding)

For all $\mathbf{G}_1\in\mathcal{F}_i$
and $(\mathcal{V}'_1,\mathcal{V}'_2)\in
\mathcal{V}(i,i,\mathcal{V}_1,\mathcal{V}_2)$, declare an error if
$|\mathcal{T}_1(\mathbf{G}_1,\mathcal{V}'_1,\mathcal{V}'_2)|<n_1(\mathbf{G}_1)/\big(\binom{K_1}{i}\binom{K_2}{i}\big)$,
otherwise each source in $\mathcal{V}'_1$ transmits
$n_1(\mathbf{G}_1)/\big(\binom{K_1}{i}\binom{K_2}{i}\big)$
information bits, which are supposed to be transmitted through
$\mathbf{G}_1$, using the time indices in
$\mathcal{T}_1(\mathbf{G}_1,\mathcal{V}'_1,\mathcal{V}'_2)$ to the
nodes in $\mathcal{V}'_2$.
\item (Relaying for $m\in\{2,\cdots,M\}$)

For all $\mathbf{G}_m\in\mathcal{F}_i$
and $(\mathcal{V}'_m,\mathcal{V}'_{m+1})\in
\mathcal{V}(i,i,\mathcal{V}_m,\mathcal{V}_{m+1})$, declare an error
if
$|\mathcal{T}_m(\mathbf{G}_m,\mathcal{V}'_m,\mathcal{V}'_{m+1})|$ is less than $n_m(\mathbf{G}_m)/\big(\binom{K_m}{i}\binom{K_{m+1}}{i}\big)$,
otherwise each node in $\mathcal{V}'_m$ transmits
$n_m(\mathbf{G}_m)/\big(\binom{K_m}{i}\binom{K_{m+1}}{i}\big)$
received bits, which are supposed to be transmitted through
$\mathbf{G}_m$, using the time indices in
$\mathcal{T}_m(\mathbf{G}_m,\mathcal{V}'_m,\mathcal{V}'_{m+1})$ to
the nodes in $\mathcal{V}'_{m+1}$. If $m=M$, the transmit bits are
constructed by the received bits that originate from
$\mathcal{S}(\mathcal{V}'_{M+1})$, where
$\mathcal{S}(\mathcal{V}'_{M+1})$ is the set of sources of
$\mathcal{V}'_{M+1}$.
\end{itemize}

From the proposed scheme, the transmission rates are given by
\begin{eqnarray}
R_1=\cdots=R_K\!\!\!\!\!\!\!\!&&=\frac{1}{Kn_B}\sum_{i=1}^{K_{\min}}i\sum_{\mathbf{G}_1\in\mathcal{F}_i}n_1(\mathbf{G}_1)\nonumber\\
&&=\frac{1}{Kn_B}\sum_{i=1}^{K_{\min}}i\sum_{\mathbf{G}_1,\cdots,\mathbf{G}_N\in\mathcal{F}_i}n_{\alpha}(\mathbf{G}_1,\cdots,\mathbf{G}_N)=R.
\end{eqnarray}
Let $s_k(i)$ denote the $i$-th information bit of the $k$-th source
and $t_{k,m}(i)$ denote the time index of the received signal
originating from $s_k(i)$ at the $m$-th hop, where
$i\in\{1,\cdots,2^{n_B R}\}$. That is, $s_k(i)$ is transmitted using
the time indices $t_{k,1}(i)$ to $t_{k,M}(i)$ during the multi-hop
transmission. The detailed decoding of the $k$-th destination is as
follows.

\begin{itemize}
\item (Decoding)

For $i\in\{1,\cdots,n_B R\}$, the $k$-th destination sets
$\hat{s}_k(i)=y_{k,M}[t_{k,M}(i)]$.
\end{itemize}

\subsubsection{Achievable rate region}
We derive the achievable rate region of the proposed scheme. Let
$E_m$ denote the event such that
\begin{equation}
|\mathcal{T}_m(\mathbf{G}_m,\mathcal{V}'_m,\mathcal{V}'_{m+1})|<\frac{n_m(\mathbf{G}_m)}{\binom{K_m}{i}\binom{K_{m+1}}{i}}
\end{equation}
for any $\mathbf{G}_m\in\mathcal{F}_i$, $(\mathcal{V}'_m,\mathcal{V}'_{m+1})\in \mathcal{V}(i,i,\mathcal{V}_m,\mathcal{V}_{m+1})$, and $i\in\{1,\cdots,K_{\min}\}$.
The following lemma shows that there is no error if $(\cup_{m=1}^M
E_m)^c$ occurs.
\begin{lemma} \label{LEM:P_e_multi}
Suppose a linear binary field relay network with $M\geq2$. If the
network has a minimum-dimensional bottleneck-hop and $p_{j,i,m}=p$
for all $i$, $j$, and $m$, then
\begin{equation}
P^{(n_B)}_{e,k}\leq\sum_{m=1}^M\Pr(E_m) \label{EQ:p_n_b}
\end{equation}
for all $k\in\{1,\cdots,K\}$.
\end{lemma}
\begin{proof}
The proof is in Appendix I.
\end{proof}

The following theorem characterizes $R$ that guarantees
$P^{(n_B)}_{e,k}\to 0$ as $n_B\to\infty$.

\begin{theorem} \label{THM:achievable_rate_multi}
Suppose a linear binary field relay network with $M\geq2$. If the
network has a minimum-dimensional bottleneck-hop and $p_{j,i,m}=p$
for all $i$, $j$, and $m$, then
\begin{equation}
R_k=\frac{1}{K}\sum_{i=1}^{K_{\min}}i\sum_{\mathbf{G}\in\mathcal{F}_i}\min\{P_{\alpha}(\mathbf{G}),P_{\beta}(\mathbf{G}^{-1})\}
\label{EQ:rate_multi}
\end{equation}
is achievable for all $k\in\{1,\cdots, K\}$, where $P_{\alpha}(\mathbf{G})$ and $P_{\beta}(\mathbf{G})$ are defined in (\ref{EQ:p_A}) and (\ref{EQ:p_B}), respectively.
\end{theorem}
\begin{proof}
Let us consider
$|\mathcal{T}_m(\mathbf{G}_m,\mathcal{V}'_m,\mathcal{V}'_{m+1})|$.
By the weak law of large numbers \cite{Csiszar:81}, there exists a
sequence $\epsilon_{n_B}\to 0$ as $n_B\to\infty$ such that the
probability
\begin{eqnarray}
&&|\mathcal{T}_m(\mathbf{G}_m,\mathcal{V}'_m,\mathcal{V}'_{m+1})|\nonumber\\
&&\geq n_B(\Pr(\mathbf{H}_{\bar{\mathcal{V}}_{m,\operatorname{tx}}[t],\bar{\mathcal{V}}_{m,\operatorname{rx}}[t]}[t]=\mathbf{G}_m,\bar{\mathcal{V}}_{m,\operatorname{tx}}[t]=\mathcal{V}_m',\bar{\mathcal{V}}_{m,\operatorname{rx}}[t]=\mathcal{V}_{m+1}')-\delta_{n_B})\nonumber\\
&& = n_B\left(P_G(\mathbf{G}_m)/\left(\binom{K_m}{i}\binom{K_{m+1}}{i}\right)-\delta_{n_B}\right)
\end{eqnarray}
for all $\mathbf{G}_m\in\mathcal{F}_i$, $(\mathcal{V}'_m,\mathcal{V}'_{m+1})\in \mathcal{V}(i,i,\mathcal{V}_m,\mathcal{V}_{m+1})$, and $i\in\{1,\cdots,K_{\min}\}$ is
greater than or equal to $1-\epsilon_{n_B}$, where $\delta_{n_B}\to
0$ as $n_B\to\infty$.
Here the equality holds from the third property of Lemma \ref{LEM:mapping1}.
This indicates that $\Pr(E_m)\leq \epsilon_{n_B}$ if
\begin{equation}
n_m(\mathbf{G}_m)\leq
n_B\left(P_G(\mathbf{G}_m)-\binom{K_m}{i}\binom{K_{m+1}}{i}\delta_{n_B}\right)
\label{EQ:condition_m}
\end{equation}
for all $\mathbf{G}_m\in\mathcal{F}_i$ and $i\in\{1,\cdots,K_{\min}\}$.
For $m\in\{1,\cdots,N\}$, from (\ref{EQ:n_m}), we also have
\begin{eqnarray}
n_m(\mathbf{G}_m)\!\!\!\!\!\!\!\!&&\leq
\sum_{\mathbf{G}_1,\cdots,\mathbf{G}_{m-1},\mathbf{G}_{m+1},\cdots,\mathbf{G}_N\in\mathcal{F}_i}c_2^{-1} n_B R \zeta_i^{-(N-1)}\prod_{l=1}^{N}P_G(\mathbf{G}_l)\nonumber\\
&&=c_2^{-1}n_BRP_G(\mathbf{G}_m),
\label{EQ:inequal_1}
\end{eqnarray}
where we use the fact that $n_{\alpha}(\mathbf{G}_1,\cdots,\mathbf{G}_N)\leq c_2^{-1}n_B R \zeta_i^{-(N-1)}\prod_{l=1}^{N}P_G(\mathbf{G}_l)$ from (\ref{eq:n_a1}).
Similarly, from (\ref{eq:n_a2}) and (\ref{EQ:n_m}), $n_m(\mathbf{G}_m)\leq c_2^{-1}n_BRP_G(\mathbf{G}_m)$ for $m\in\{N+1,\cdots,M\}$.
Then, the condition in (\ref{EQ:condition_m}) can be satisfied if
\begin{equation}
R\leq\frac{c_2\big(P_G(\mathbf{G}_m)-\binom{K_m}{i}\binom{K_{m+1}}{i}\delta_{n_B}\big)}{P_G(\mathbf{G}_m)}
\label{EQ:condition_m_2}
\end{equation}
for all $\mathbf{G}_m\in\mathcal{F}_i$ and $i\in\{1,\cdots,K_{\min}\}$.
Hence, we set $R=c_2(1-\delta^*_{n_B})$, where
$\delta^*_{n_B}=\frac{(K_{\max}!)^2\delta_{n_B}}{\min_{i\in\{1,\cdots,K_{\min}\},\mathbf{G}'\in\mathcal{F}_i}\{P_G(\mathbf{G}')\}}$, which converges
to zero as $n_B\to\infty$.
Therefore, from Lemma
\ref{LEM:P_e_multi}, we have $P^{(n_B)}_{e,k}\leq
M\epsilon_{n_B}$, which converges
to zero as $n_B\to\infty$.
In conclusion,
\begin{eqnarray}
R_k=c_2=\frac{1}{K}\sum_{i=1}^{K_{\min}}i\sum_{\mathbf{G}\in\mathcal{F}_i}\min\{P_{\alpha}(\mathbf{G}),P_{\beta}(\mathbf{G}^{-1})\}
\end{eqnarray}
is achievable for all $k\in\{1,\cdots,K\}$, which completes the proof.
\end{proof}

For $M=2$, the proposed scheme pairs $\mathbf{G}_1$ and $\mathbf{G}_2$ satisfying $\mathbf{G}_2=\mathbf{G}_1^{-1}$ and the achievable rate in (\ref{EQ:rate_multi}) is given by
\begin{equation}
R_k=\frac{1}{K}\sum_{i=1}^{K_{\min}}i\sum_{\mathbf{G}\in\mathcal{F}_i}\min\left\{P_G(\mathbf{G}),P_G(\mathbf{G}^{-1})\right\}.
\label{EQ:rate_even}
\end{equation}
Let us now consider the capacity achieving case.
The following corollary shows that if $p=1/2$, the sum capacity is given by the average rank of the channel matrix of the bottleneck-hop.

\begin{corollary} \label{CO:capacity_multi}
Suppose a linear binary field relay network with $M\geq2$. If the
network has a minimum-dimensional bottleneck-hop and $p_{j,i,m}=1/2$
for all $i$, $j$, and $m$, the sum capacity is given by
\begin{equation}
C_{\operatorname{sum}}=2^{-K_{m_0+1}K_{m_0}}\!\!\!\!\!\sum_{\mathbf{H}\in\mathbb{F}_2^{K_{m_0+1}\times
K_{m_0}}}\rank(\mathbf{H}). \label{EQ:C_sum2}
\end{equation}
\end{corollary}
\begin{proof}
From (\ref{EQ:p_G_equal_prob}), we have
\begin{equation}
\zeta_i=\sum_{\mathbf{G}\in\mathcal{F}_i}2^{-K_{m_0+1}K_{m_0}}\frac{N_{K_{m_0+1},K_{m_0}}(i)}{N_{i,i}(i)}=2^{-K_{m_0+1}K_{m_0}}N_{K_{m_0+1},K_{m_0}}(i)
\end{equation}
and
\begin{eqnarray}
P_{\alpha}(\mathbf{G})\!\!\!\!\!\!\!\!&&=\zeta_i^{-(N-1)}\sum_{\underset{\mathbf{G}_{N}\cdots\mathbf{G}_1=\mathbf{G}}{\mathbf{G}_1,\cdots,\mathbf{G}_N\in \mathcal{F}_i,}}\prod_{m=1}^{N}2^{-K_{m_0+1}K_{m_0}}\frac{N_{K_{m_0+1},K_{m_0}}(i)}{N_{i,i}(i)}\nonumber\\
&&=2^{-K_{m_0+1}K_{m_0}}\frac{N_{K_{m_0+1},K_{m_0}}(i)}{(N_{i,i}(i))^N}\sum_{\underset{\mathbf{G}_{N}\cdots\mathbf{G}_1=\mathbf{G}}{\mathbf{G}_1,\cdots,\mathbf{G}_{N}\in\mathcal{F}_i}}1\nonumber\\
&&=2^{-K_{m_0+1}K_{m_0}}\frac{N_{K_{m_0+1},K_{m_0}}(i)}{N_{i,i}(i)}.
\end{eqnarray}
Similarly, $P_{\beta}(\mathbf{G})=2^{-K_{m_0+1}K_{m_0}}\frac{N_{K_{m_0+1},K_{m_0}}(i)}{N_{i,i}(i)}$.
Then, from (\ref{EQ:rate_multi}),
\begin{eqnarray}
R_k\!\!\!\!\!\!\!\!\!&&=\frac{1}{K}2^{-K_{m_0+1}K_{m_0}}\sum_{i=1}^{K_{\min}}iN_{K_{m_0+1},K_{m_0}}(i)\nonumber\\
&&=\frac{1}{K}2^{-K_{m_0+1}K_{m_0}}\sum_{\mathbf{H}\in\mathbb{F}_2^{K_{m_0+1}\times
K_{m_0}}}\rank(\mathbf{H})
\end{eqnarray}
is achievable for all $k\in\{1,\cdots,K\}$. Hence the sum rate in
(\ref{EQ:C_sum2}) is achievable, which coincides with the sum rate
upper bound in (\ref{EQ:converse_multi}). In conclusion, Corollary
\ref{CO:capacity_multi} holds.
\end{proof}
\begin{remark} \label{RE:q_ary_multi}
Corollary \ref{CO:capacity_multi} can be directly extended to a
general linear finite field relay network in which inputs, outputs,
and channels are in $\mathbb{F}_q$ and channels are i.i.d. uniformly
distributed over $\mathbb{F}_q$. Then
\begin{equation}
C_{\operatorname{sum}}=q^{-K_{m_0+1}K_{m_0}}\sum_{\mathbf{H}\in\mathbb{F}_q^{K_{m_0+1}\times
K_{m_0}}}\rank(\mathbf{H})\log q.
\end{equation}
\end{remark}

Notice that Corollary \ref{CO:capacity_multi} shows that the sum
rate of $\mathbb{E}(\rank(\mathbf{H}_{m_0}[1]))$ is achievable,
which is the \emph{multi-input multi-output (MIMO) capacity of the
bottleneck-hop}. This result also shows that the \emph{max-flow
min-cut theorem} holds for a certain class of channel distributions
and network topologies.


For $2$--$2$--$2$ networks, we characterize the sum capacity for
more general classes of channel distributions by applying
deterministic channel pairing.

\begin{theorem} \label{THM:2_2_2networks}
Suppose a linear binary field relay network with $M=2$ and
$K_1=K_2=K_3=2$.
\begin{enumerate}
\item For a symmetric channel satisfying $p_{1,1,1}=p_{2,2,1}=p_{1,1,2}=p_{2,2,2}$ and $p_{1,2,1}=p_{2,1,1}=p_{1,2,2}=p_{2,1,2}$ or a $Z$ channel satisfying $p_{2,1,1}=p_{2,1,2}=0$, $p_{1,1,1}=p_{1,1,2}$, $p_{1,2,1}=p_{1,2,2}$, and $p_{2,2,1}=p_{2,2,2}$, the sum capacity is given by
\begin{equation}
C_{\operatorname{sum}}=\mathbb{E}(\operatorname{rank}(\mathbf{H}_1[1])).
\end{equation}
\item For a $Z$ channel satisfying $p_{2,1,1}=p_{2,1,2}=0$, $p_{1,1,1}=p_{2,2,2}$, $p_{1,2,1}=p_{1,2,2}$, and $p_{2,2,1}=p_{1,1,2}$, the sum capacity is given by
\begin{eqnarray}
C_{\operatorname{sum}}=\min\left\{2p_{1,1,1},2p_{2,2,1}\right\}.
\end{eqnarray}
\end{enumerate}
\end{theorem}
\begin{proof}
We use deterministic channel pairing between the first and the
second hops. The overall block encoding and relaying structure making such pairing
possible is the same as in the previous scheme.

\begin{figure}[t!]
  \begin{center}
  \scalebox{1.1}{\includegraphics{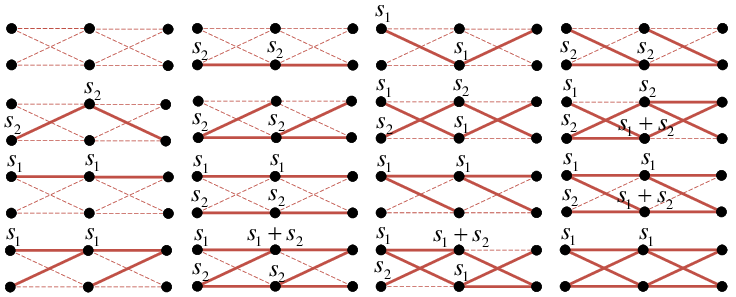}}
  \caption{Deterministic channel pairing between the first and the second hops.}
  \label{FIG:deterministic_mapping_2hop}
  \end{center}
\end{figure}

Let us prove the first result. Fig.
\ref{FIG:deterministic_mapping_2hop} illustrates the deterministic
channel pairing between the first and the second hops and related
encoding and relaying. The solid lines and the dashed lines denote
the corresponding channels are ones and zeros, respectively. The
symbols in the figure denote the transmit signals of the nodes and
the nodes with no symbol transmit zeros, where $s_k$ denotes the
information bit of the $k$-th source. Let $p^{(1)}_m$ to
$p_m^{(16)}$ denote $16$ possible instances of $\mathbf{H}_m[t]$ as
shown in Fig. \ref{FIG:instances}, where $m\in\{1,2\}$.
Then the achievable sum rate of the deterministic pairing in Fig.
\ref{FIG:deterministic_mapping_2hop} is given by
\begin{eqnarray}
\!\!\!\!\!\!\!\!R_{\operatorname{sum}}\!\!\!\!\!\!\!\!\!&&=\sum_{i\in\{2,4,6,9,11,13,16\}}\min\{p_1^{(i)},p_2^{(i)}\}+\min\{p_1^{(3)},p_2^{(5)}\}+\min\{p_1^{(5)},p_2^{(3)}\}\nonumber\\
\!\!\!\!\!\!\!\!&&{~~}+2\sum_{i\in\{7,10,12,14\}}\min\{p_1^{(i)},p_2^{(i)}\}+2\min\{p_1^{(8)},p_2^{(15)}\}+2\min\{p_1^{(15)},p_2^{(8)}\}.
\label{EQ:achievable_rate_2_2_2}
\end{eqnarray}

\begin{figure}[t!]
  \begin{center}
  \scalebox{1}{\includegraphics{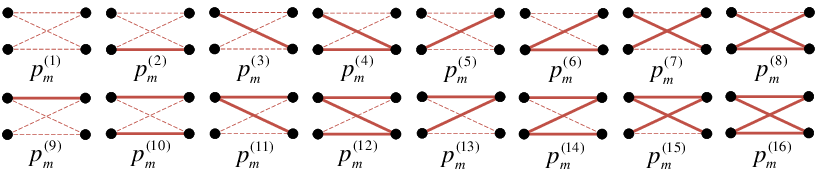}}
  \caption{$16$ possible instances of $\mathbf{H}_m[t]$.}
  \label{FIG:instances}
  \end{center}
\end{figure}

Since the probabilities of each paired $\mathbf{H}_1[t_1]$ and
$\mathbf{H}_2[t_2]$ are the same, from
(\ref{EQ:achievable_rate_2_2_2}), we have
\begin{eqnarray}
R_{\operatorname{sum}}=\sum_{i\in\{2,3,4,5,6,9,11,13,16\}}p_1^{(i)}+2\sum_{i\in\{7,8,10,12,14,15\}}p_1^{(i)}=\mathbb{E}(\operatorname{rank}(\mathbf{H}_1[1])).
\end{eqnarray}
Notice that the achievable sum rate coincides with the sum rate
upper bound in (\ref{EQ:converse_multi}).

Now let us prove the second result. Unlike the previous case,
the probabilities of some paired $\mathbf{H}_1[t_1]$ and
$\mathbf{H}_2[t_2]$ in Fig. \ref{FIG:deterministic_mapping_2hop} are
not the same. Let us denote $p_a=p_{1,1,1}=p_{2,2,2}$,
$p_b=p_{1,2,1}=p_{1,2,2}$, and $p_c=p_{2,2,1}=p_{1,1,2}$. For
$p_a\geq p_c$, from (\ref{EQ:achievable_rate_2_2_2}),
\begin{eqnarray}
R_{\operatorname{sum}}\!\!\!\!\!\!\!\!\!&&=p_{1}^{(2)}+p_{1}^{(6)}+p_{2}^{(9)}+p_{2}^{(13)}+2p_{1}^{(10)}+2p_{1}^{(14)}\nonumber\\
&&=(1-p_a)(1-p_b)p_c+(1-p_a)p_bp_c+(1-p_a)(1-p_b)p_c+(1-p_a)p_bp_c\nonumber\\
&&{~~}+2p_a(1-p_b)p_c+2p_ap_bp_c=2p_c \label{EQ:sum_rate_222_1}
\end{eqnarray}
is achievable. By setting $\Omega_1=\{v_{1,1},v_{1,2},v_{2,1}\}$, we
have
\begin{equation}
R_{\operatorname{sum}}\leq\mathbb{E}(\rank(\left[[h_{2,2,1}[1],0]^T,[0,h_{1,1,2}[1]]^T)\right]^T)=2p_c,
\end{equation}
which coincides with (\ref{EQ:sum_rate_222_1}). Similarly, for $p_a<
p_c$, from (\ref{EQ:achievable_rate_2_2_2}),
\begin{equation}
R_{\operatorname{sum}}=p_{2}^{(2)}+p_{2}^{(6)}+p_{1}^{(9)}+p_{1}^{(13)}+2p_{1}^{(10)}+2p_{1}^{(14)}=2p_a
\label{EQ:sum_rate_222_2}
\end{equation}
is achievable. From $\Omega_2=\{v_{1,1}\}$ and
$\Omega_3=\{v_{1,1},v_{1,2},v_{1,3},v_{2,1},v_{2,2}\}$, we have
$R_1\leq p_a$ and $R_2\leq p_a$, respectively. Then
$R_{\operatorname{sum}}\leq2p_a$, which coincides with
(\ref{EQ:sum_rate_222_2}). In conclusion, Theorem
\ref{THM:2_2_2networks} holds.
\end{proof}

\begin{figure}[t!]
  \begin{center}
  \scalebox{0.8}{\includegraphics{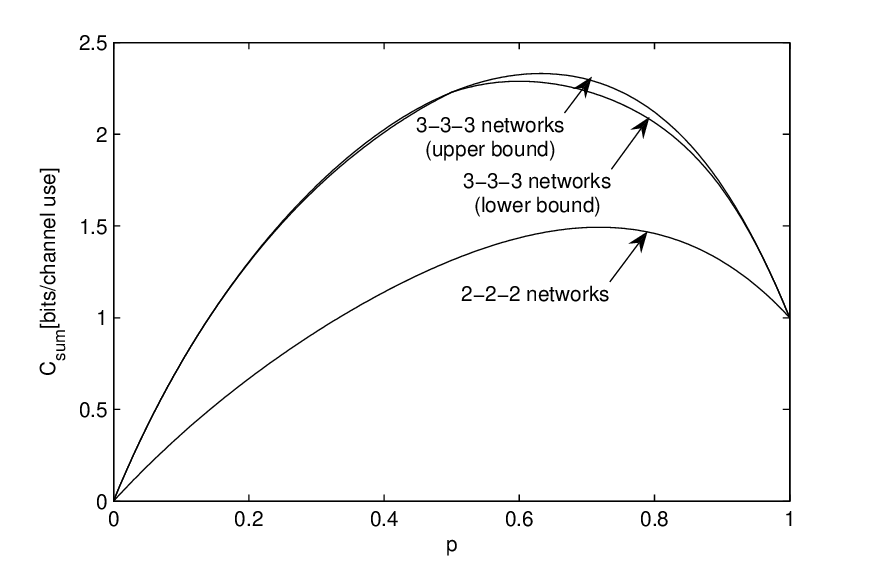}}
  \caption{Sum capacity when $p_{j,i,m}=p$ for all $i$, $j$, and $m$.}
  \label{FIG:achievable}
  \end{center}
\end{figure}

\begin{example} [$2$--$2$--$2$ and $3$--$3$--$3$ networks]
Fig. \ref{FIG:achievable} plots sum rates of two-hop networks with
$p_{j,i,m}=p$. For $2$--$2$--$2$ networks, the sum capacity is given
by $C_{\operatorname{sum}}= 4pq^3+8p^2q^2+8p^3q+p^4$, where $q=1-p$.
Notice that the considered channel distribution is a special case of
the symmetric channel in Theorem \ref{THM:2_2_2networks}. Therefore,
we can characterize the sum capacity for all $p\in[0,1]$. For
$3$--$3$--$3$ networks, we obtain $C_{\operatorname{sum}}\geq
9pq^8+54p^2q^7+168p^3q^6+279p^4q^5+216p^5q^4+72|p^5q^4-p^6q^3|+216\min\{p^5q^4,p^6q^3\}+90p^6q^3+90p^7q^2+18p^8q+p^9$
and $C_{\operatorname{sum}}\leq
9pq^8+54p^2q^7+168p^3q^6+279p^4q^5+324p^5q^4+198p^6q^3+90p^7q^2+18p^8q+p^9$.
The lower and upper bounds are the same when $p=\frac{1}{2}$, which
coincides with the result of Corollary \ref{CO:capacity_multi} (if
$p=0$ or $1$ the lower and upper bounds are trivially the same).
\end{example}

\begin{example} [Networks with $K=K_1=\cdots=K_{M+1}$]
Suppose a linear finite field relay network with
$K=K_1=\cdots=K_{M+1}$ in which inputs, outputs, and channels are in
$\mathbb{F}_q$ and channels are i.i.d. uniformly distributed over
$\mathbb{F}_q$. From Remarks 1 and 2, we have
\begin{equation}
C_{\operatorname{sum}}=\begin{cases}\frac{K}{2}\log q &\mbox{ if }M=1,\\
\mathbb{E}(\rank(\mathbf{H}_{m_0}[1]))\log q&\mbox{ if }M\geq2.
\end{cases}
\end{equation}
For $K=K_1=\cdots=K_{M+1}=2$ and $q=2$, $C_{\operatorname{sum}}$ is
given by $1$ if $M=1$ and $21/16$ if $M\geq 2$.
\end{example}

\section{Conclusion} \label{sec:conclusion}
In this paper, we studied layered linear binary field relay networks
with time-varying channels, which exhibit broadcast, interference,
and fading natures of wireless communications. Capacity
characterization of such relay networks with multiple S--D pairs is
quite challenging because the transmission of other session acts as
inter-user interference.
We observed that the fading can play an
important role in mitigating interference that leads to the capacity
characterization for some classes of channel distributions and
network topologies. For these classes, we showed that the capacity
region of single-hop networks and the sum capacity of multi-hop
networks can be interpreted as the max-flow min-cut theorem.

\section*{Appendix I\\Upper Bound on the Probability of Error}
\emph{{~~}Proof of Lemma \ref{LEM:P_e_single}:} Let us assume that
($E_1\cup E_2)^c$ occurs. Then, from the assumption, each
source can transmit $n(\mathbf{H}_1)$ bits using the time indices in
$\mathcal{T}_1(\mathbf{H}_1)$ for all $\mathbf{H}_1$. Since
$n(\mathbf{H}_1)=n(\mathbf{H}_1+\mathbf{I})$, from the assumption,
each source can retransmit all information bits that were
transmitted during $\mathcal{T}_1(\mathbf{H}_1+\mathbf{I})$
using the time indices in $\mathcal{T}_2(\mathbf{H}_1)$ for all $\mathbf{H}_1$. Lastly,
there is no decoding error if $(E_1\cup E_2)^c$ occurs since
$\mathbf{H}_1[t_1(i)]+\mathbf{H}_1[t_2(i)]=\mathbf{I}$, meaning
$\hat{s}_k(i)=s_k(i)$. In conclusion, from the union bound, we
obtain $P^{(n)}_{e,k}\leq \Pr(E_1)+\Pr(E_2)$, which
completes the proof. \hfill$\blacksquare$

\emph{{~~}Proof of Lemma \ref{LEM:P_e_multi}:} Let us assume that
$(\cup_{m=1}^M {E_m})^c$ occurs. Then each source can transmit all information bits to the nodes in the next layer.
Consider the $m$-th hop transmission through
$\mathbf{G}_m\in\mathcal{F}_i$, where $m\in\{2,\cdots,M-1\}$. 
Each node in $\mathcal{V}'_m$ receives
$n_m(\mathbf{G}_m)/\big(\binom{K_{m-1}}{i}\binom{K_m}{i}\big)$ bits from $\mathcal{V}'_{m-1}$
that should be transmitted through $\mathbf{G}_m$. Since there are
$\binom{K_{m-1}}{i}$ candidates for $\mathcal{V}'_{m-1}$, a total of
$n_m(\mathbf{G}_m)/\binom{K_m}{i}$ bits should be transmitted
through $\mathbf{G}_m$. From the assumption, each node in
$\mathcal{V}'_m$ is able to transmit
$n_m(\mathbf{G}_m)/\big(\binom{K_m}{i}\binom{K_{m+1}}{i}\big)$ bits
to the nodes in $\mathcal{V}'_{m+1}$ using the time indices in
$\mathcal{T}_m(\mathbf{G}_m,\mathcal{V}'_m,\mathcal{V}'_{m+1})$.
Since there are $\binom{K_{m+1}}{i}$ candidates for $\mathcal{V}'_{m+1}$, each node in $\mathcal{V}'_m$ can transmit a total of $n_m(\mathbf{G}_m)/\binom{K_m}{i}$ bits through $\mathbf{G}_m$.
Hence, each node in $\mathcal{V}'_m$ can transmit all received bits.
Consider the last hop transmission. Similar to the previous hops,
each node in $\mathcal{V}'_M$ receives
$n_M(\mathbf{G}_M)/\binom{K_{M}}{i}$ bits that should be transmitted
through $\mathbf{G}_M$ and, among them,
$n_M(\mathbf{G}_M)/\big(\binom{K_1}{i}\binom{K_{M}}{i}\big)$ bits
are originated from $\mathcal{S}(\mathcal{V}'_{M+1})$. From the
assumption, each node in $\mathcal{V}'_M$ is able to transmit
$n_M(\mathbf{G}_M)/\big(\binom{K_M}{i}\binom{K_{M+1}}{i}\big)$ bits
to the nodes in $\mathcal{V}'_{M+1}$ using the time indices in
$\mathcal{T}_M(\mathbf{G}_M,\mathcal{V}'_M,\mathcal{V}'_{M+1})$.
Hence, each node in $\mathcal{V}'_M$ can transmit all received bits
because
$n_M(\mathbf{G}_M)/\big(\binom{K_1}{i}\binom{K_{M}}{i}\big)$ is equal to $n_M(\mathbf{G}_M)/\big(\binom{K_M}{i}\binom{K_{M+1}}{i}\big)$,
where we use the fact that $K=K_1=K_{M+1}$.

\begin{table}
\caption{Notations used in Appendix I.}
\label{Table:abbreviations}
\begin{equation*}
  \begin{array}{|c|c|}
  \hline
  P^{(1)}_m(\mathbf{H}_m) & \Pr(\mathbf{H}_m[t]=\mathbf{H}_m)\\
  \hline
  P^{(2)}_m(\mathbf{H}) & \Pr(\mathbf{H}_{\mathcal{V}_{m,\operatorname{tx}}[t],\mathcal{V}_{m,\operatorname{rx}}[t]}[t]=\mathbf{H})\\
  \hline
  P^{(3)}_m(\mathcal{V}',\mathcal{V}''\big|\mathbf{H}_m) & \Pr(\mathcal{V}_{m,\operatorname{tx}}[t]=\mathcal{V}',\mathcal{V}_{m,\operatorname{rx}}[t]=\mathcal{V}''\big|\mathbf{H}_m[t]=\mathbf{H}_m)\\
  \hline
  P^{(4)}_m(\mathbf{H}\big|\mathbf{H}_m,\mathcal{V}',\mathcal{V}'')& \Pr(\mathbf{H}_{\mathcal{V}_{m,\operatorname{tx}}[t],\mathcal{V}_{m,\operatorname{rx}}[t]}[t]=\mathbf{H}\big|\mathbf{H}_m[t]=\mathbf{H}_m,\mathcal{V}_{m,\operatorname{tx}}[t]=\mathcal{V}',\mathcal{V}_{m,\operatorname{rx}}[t]=\mathcal{V}'')\\
  \hline
  P^{(5)}_m(\mathcal{V}',\mathcal{V}'') & \Pr(\mathcal{V}_{m,\operatorname{tx}}[t]=\mathcal{V}',\mathcal{V}_{m,\operatorname{rx}}[t]=\mathcal{V}'')\\
  \hline
  P^{(6)}_m(\mathbf{G}) & \Pr(\mathbf{H}_{\bar{\mathcal{V}}_{m,\operatorname{tx}}[t],\bar{\mathcal{V}}_{m,\operatorname{rx}}[t]}[t]=\mathbf{G})\\
  \hline
  P^{(7)}_m(\mathcal{V}',\mathcal{V}''\big|\mathbf{H}) & \Pr(\bar{\mathcal{V}}_{m,\operatorname{tx}}[t]=\mathcal{V}',\bar{\mathcal{V}}_{m,\operatorname{rx}}[t]=\mathcal{V}''\big|\mathbf{H}_{\mathcal{V}_{m,\operatorname{tx}}[t],\mathcal{V}_{m,\operatorname{rx}}[t]}[t]=\mathbf{H})\\
  \hline
  P^{(8)}_m(\mathbf{G}\big|\mathbf{H},\mathcal{V}',\mathcal{V}'')& \Pr(\mathbf{H}_{\bar{\mathcal{V}}_{m,\operatorname{tx}}[t],\bar{\mathcal{V}}_{m,\operatorname{rx}}[t]}[t]=\mathbf{G}\big|\mathbf{H}_{\mathcal{V}_{m,\operatorname{tx}}[t],\mathcal{V}_{m,\operatorname{rx}}[t]}[t]=\mathbf{H},\bar{\mathcal{V}}_{m,\operatorname{tx}}[t]=\mathcal{V}',\bar{\mathcal{V}}_{m,\operatorname{rx}}[t]=\mathcal{V}'')\\
  \hline
  P^{(9)}_m(\mathbf{G},\mathcal{V}',\mathcal{V}'') & \Pr(\mathbf{H}_{\bar{\mathcal{V}}_{m,\operatorname{tx}}[t],\bar{\mathcal{V}}_{m,\operatorname{rx}}[t]}[t]=\mathbf{G},\bar{\mathcal{V}}_{m,\operatorname{tx}}[t]=\mathcal{V}',\bar{\mathcal{V}}_{m,\operatorname{rx}}[t]=\mathcal{V}'')\\
  \hline
  \end{array}
\end{equation*}
\end{table}

Lastly, consider the estimated bit $\hat{s}_k(i)$ at the $k$-th
destination. Since the overall channel matrix from
$\bar{\mathcal{V}}_{\operatorname{tx},1}[t_{k,1}(i)]$ to
$\bar{\mathcal{V}}_{\operatorname{rx},M}[t_{k,M}(i)]$ is given by
\begin{equation}
\mathbf{H}_{\bar{\mathcal{V}}_{\operatorname{tx},M}[t_{k,M}(i)],\bar{\mathcal{V}}_{\operatorname{rx},M}[t_{k,M}(i)]}[t_{k,M}(i)]\cdots
\mathbf{H}_{\bar{\mathcal{V}}_{\operatorname{tx},1}[t_{k,1}(i)],\bar{\mathcal{V}}_{\operatorname{rx},1}[t_{k,1}(i)]}[t_{k,1}(i)]=\mathbf{I},
\end{equation}
we obtain $\hat{s}_k(i)=s_k(i)$. Hence, there is no error if
$(\cup_{m=1}^M {E_m})^c$ occurs. In conclusion, from the union
bound, we obtain $P^{(n_B)}_{e,k}\leq\sum_{m=1}^M\Pr(E_m)$, which
completes the proof. \hfill$\blacksquare$

\section*{Appendix II\\Probability Distributions of Sub-channel Matrices} \label{APP:app1}
In this appendix, we prove the probability distributions shown in Lemma \ref{LEM:mapping1}. For
notational simplicity, we will use the shorthand notations in Table
\ref{Table:abbreviations}.

\emph{{~~}Proof of Lemma \ref{LEM:mapping1}.(1):} We assume that
$|\mathcal{V}_{m,\operatorname{tx}}[t]|=K_{m_0}$ and
$|\mathcal{V}_{m,\operatorname{rx}}[t]|=K_{m_0+1}$ in the proof. But
the same result holds for the case where
$|\mathcal{V}_{m,\operatorname{tx}}[t]|=K_{m_0+1}$ and
$|\mathcal{V}_{m,\operatorname{rx}}[t]|=K_{m_0}$. We have
\begin{eqnarray}
P^{(2)}_m\left(\mathbf{H}\right)\!\!\!\!\!\!\!\!&&=\sum_{\underset{\mathcal{V}(K_{m_0},K_{m_0+1},\mathcal{V}_{m},\mathcal{V}_{m+1})}{(\mathcal{V}',\mathcal{V}'')\in}}\sum_{\mathbf{H}_m\in \mathbb{F}_2^{K_{m+1}\times K_m}}P^{(1)}_m(\mathbf{H}_m)P^{(3)}_m\left(\mathcal{V}',\mathcal{V}''\big|\mathbf{H}_m\right)P_m^{(4)}\left(\mathbf{H}\big|\mathbf{H}_m,\mathcal{V}',\mathcal{V}''\right)\nonumber\\
&&\overset{(a)}{=}\sum_{\underset{\mathcal{V}(K_{m_0},K_{m_0+1},\mathcal{V}_{m},\mathcal{V}_{m+1})}{(\mathcal{V}',\mathcal{V}'')\in}}P_m^{(5)}\left(\mathcal{V}',\mathcal{V}''\right)\sum_{\mathbf{H}_m\in \mathbb{F}_2^{K_{m+1}\times K_m}}P_m^{(1)}(\mathbf{H}_m)P_m^{(4)}\left(\mathbf{H}\big|\mathbf{H}_m,\mathcal{V}',\mathcal{V}''\right)\nonumber\\
&&\overset{(b)}{=}\sum_{\underset{\mathcal{V}(K_{m_0},K_{m_0+1},\mathcal{V}_{m},\mathcal{V}_{m+1})}{(\mathcal{V}',\mathcal{V}'')\in}}P_m^{(5)}\left(\mathcal{V}',\mathcal{V}''\right)\sum_{\mathbf{H}_m\in \mathcal{H}_{\mathcal{V}_m,\mathcal{V}_{m+1}}(\mathbf{H},\mathcal{V}',\mathcal{V}'')}P_m^{(1)}(\mathbf{H}_m)\nonumber\\
&&\overset{(c)}{=}p^u(1-p)^{K_{m_0+1}K_{m_0}-u},
\end{eqnarray}
where $(a)$ holds from the fact that
$P_m^{(3)}\left(\mathcal{V}',\mathcal{V}''\big|\mathbf{H}_m\right)=P_m^{(5)}\left(\mathcal{V}',\mathcal{V}''\right)$
because $\mathcal{V}_{m,\operatorname{tx}}[t]$ and
$\mathcal{V}_{m,\operatorname{rx}}[t]$ are chosen regardless of
channel instances, $(b)$ holds since
\begin{equation}
P_m^{(4)}\left(\mathbf{H}\big|\mathbf{H}_m,\mathcal{V}',\mathcal{V}''\right)=\begin{cases}1&\text{if } \mathbf{H}_m\in \mathcal{H}_{\mathcal{V}_m,\mathcal{V}_{m+1}}(\mathbf{H},\mathcal{V}',\mathcal{V}'')\\
0&\text{otherwise},
\end{cases}
\end{equation}
and $(c)$ holds since $\sum_{\mathbf{H}_m\in
\mathcal{H}_{\mathcal{V}_m,\mathcal{V}_{m+1}}(\mathbf{H},\mathcal{V}',\mathcal{V}'')}P_m^{(1)}(\mathbf{H}_m)=p^u(1-p)^{K_{m_0+1}K_{m_0}-u}$.
Therefore, Lemma \ref{LEM:mapping1}.(1) holds. \hfill$\blacksquare$

\emph{{~~}Proof of Lemma \ref{LEM:mapping1}.(2):} We again assume
that $|\mathcal{V}_{m,\operatorname{tx}}[t]|=K_{m_0}$ and
$|\mathcal{V}_{m,\operatorname{rx}}[t]|=K_{m_0+1}$ in the proof.
We have
\begin{eqnarray}
P_m^{(6)}(\mathbf{G})\!\!\!\!\!\!\!\!&&=\sum_{\underset{\mathcal{V}(i,i,\mathcal{V}_{m,\operatorname{tx}}[t],\mathcal{V}_{m,\operatorname{rx}}[t])}{(\mathcal{V}',\mathcal{V}'')\in}}\sum_{\mathbf{H}\in \mathbb{F}_2^{K_{m_0+1}\times K_{m_0}}}P_m^{(2)}(\mathbf{H})P_m^{(7)}(\mathcal{V}',\mathcal{V}''|\mathbf{H})P_m^{(8)}(\mathbf{G}|\mathbf{H},\mathcal{V}',\mathcal{V}'')\nonumber\\
&&\overset{(a)}{=}\sum_{\underset{\mathcal{V}(i,i,\mathcal{V}_{m,\operatorname{tx}}[t],\mathcal{V}_{m,\operatorname{rx}}[t])}{(\mathcal{V}',\mathcal{V}'')\in}}\sum_{\mathbf{H}\in\mathcal{H}^F_{\mathcal{V}_{m,\operatorname{tx}}[t],\mathcal{V}_{m,\operatorname{rx}}[t]}(\mathbf{G},\mathcal{V}',\mathcal{V}'')}P_m^{(2)}(\mathbf{H})P_m^{(7)}(\mathcal{V}',\mathcal{V}''|\mathbf{H})\nonumber\\
&&\overset{(b)}{=}\sum_{\underset{\mathcal{V}(i,i,\mathcal{V}_{m,\operatorname{tx}}[t],\mathcal{V}_{m,\operatorname{rx}}[t])}{(\mathcal{V}',\mathcal{V}'')\in}}\sum_{\mathbf{H}\in\mathcal{H}^F_{\mathcal{V}_{m,\operatorname{tx}}[t],\mathcal{V}_{m,\operatorname{rx}}[t]}(\mathbf{G},\mathcal{V}',\mathcal{V}'')}\frac{P_m^{(2)}(\mathbf{H})}{|\mathcal{V}(\mathbf{H},\mathcal{V}_{m,\operatorname{tx}}[t],\mathcal{V}_{m,\operatorname{rx}}[t])|}\nonumber\\
&&\overset{(c)}{=}\sum_{\underset{\mathcal{V}(i,i,\mathcal{V}_{m_0},\mathcal{V}_{m_0+1})}{(\mathcal{V}',\mathcal{V}'')\in}}\sum_{\mathbf{H}\in\mathcal{H}^F_{\mathcal{V}_{m_0},\mathcal{V}_{m_0+1}}(\mathbf{G},\mathcal{V}',\mathcal{V}'')}\frac{P_m^{(2)}(\mathbf{H})}{|\mathcal{V}(\mathbf{H},\mathcal{V}_{m_0},\mathcal{V}_{m_0+1})|},
\label{EQ:p_G_proof}
\end{eqnarray}
where $(a)$ holds since
\begin{equation}
P_m^{(8)}(\mathbf{G}|\mathbf{H},\mathcal{V}',\mathcal{V}'')=\begin{cases}1&\text{if } \mathbf{H}\in\mathcal{H}^F_{\mathcal{V}_{m,\operatorname{tx}}[t],\mathcal{V}_{m,\operatorname{rx}}[t]}(\mathbf{G},\mathcal{V}',\mathcal{V}'')\\
0&\text{otherwise},
\end{cases}
\end{equation}
$(b)$ holds since
$P_m^{(7)}(\mathcal{V}',\mathcal{V}''|\mathbf{H})=\frac{1}{|\mathcal{V}(\mathbf{H},\mathcal{V}_{m,\operatorname{tx}}[t],\mathcal{V}_{m,\operatorname{rx}}[t])|}$
if
$\mathbf{H}\in\mathcal{H}^F_{\mathcal{V}_{m,\operatorname{tx}}[t],\mathcal{V}_{m,\operatorname{rx}}[t]}(\mathbf{G},\mathcal{V}',\mathcal{V}'')$,
and $(c)$ holds from the facts that $P_m^{(2)}(\mathbf{H})$ is the
same for all $m$, which is the result of Lemma
\ref{LEM:mapping1}.(1), and
$|\mathcal{V}(\mathbf{H},\mathcal{V}_{m,\operatorname{tx}}[t],\mathcal{V}_{m,\operatorname{rx}}[t])|$
is the same for all $m$.

Now consider the case $p_{j,i,m}=1/2$. Since
$\rank(\mathbf{H}_{\bar{\mathcal{V}}_{m,\operatorname{tx}}[t],\bar{\mathcal{V}}_{m,\operatorname{tx}}[t]}[t])=\rank(\mathbf{H}_{\mathcal{V}_{m,\operatorname{tx}}[t],\mathcal{V}_{m,\operatorname{tx}}[t]}[t])$,
we obtain
\begin{equation}
\sum_{\mathbf{G}'\in\mathcal{F}_i}P_m^{(6)}(\mathbf{G}')=\sum_{\mathbf{H}\in\mathbb{F}^{K_{m_0+1}\times
K_{m_0}}_2,\rank(\mathbf{H})=i}P_m^{(2)}(\mathbf{H}),
\label{EQ:p_G_p_H}
\end{equation}
where
\begin{equation}
P^{(2)}_m(\mathbf{H})=2^{-K_{m_0+1}K_{m_0}} \label{EQ:ph}
\end{equation}
and
\begin{equation}
P_m^{(6)}(\mathbf{G}')=2^{-K_{m_0+1}K_{m_0}}\sum_{\underset{\mathcal{V}(\rank(\mathbf{G}'),\rank(\mathbf{G}'),\mathcal{V}_{m_0},\mathcal{V}_{m_0+1})}{(\mathcal{V}',\mathcal{V}'')\in}}\sum_{\mathbf{H}\in\mathcal{H}^F_{\mathcal{V}_{m_0},\mathcal{V}_{m_0+1}}(\mathbf{G}',\mathcal{V}',\mathcal{V}'')}\frac{1}{|\mathcal{V}(\mathbf{H},\mathcal{V}_{m_0},\mathcal{V}_{m_0+1})|}.
\label{EQ:pg}
\end{equation}
Here, (\ref{EQ:ph}) and (\ref{EQ:pg}) can be derived from Lemma
\ref{LEM:mapping1}.(1) and (\ref{EQ:p_G_proof}).
Then we will prove the following two properties:
\begin{enumerate}
\item $\sum_{\mathbf{H}\in\mathcal{H}^F_{\mathcal{V}_{m_0},\mathcal{V}_{m_0+1}}(\mathbf{G}',\mathcal{V}',\mathcal{V}'')}\frac{1}{|\mathcal{V}(\mathbf{H},\mathcal{V}_{m_0},\mathcal{V}_{m_0+1})|}$ is the same for all $\mathcal{V}'$ and $\mathcal{V}''$.
\item $\sum_{\mathbf{H}\in\mathcal{H}^F_{\mathcal{V}_{m_0},\mathcal{V}_{m_0+1}}(\mathbf{G}',\mathcal{V}',\mathcal{V}'')}\frac{1}{|\mathcal{V}(\mathbf{H},\mathcal{V}_{m_0},\mathcal{V}_{m_0+1})|}$ is the same for all $\mathbf{G}'$ having the same rank.
\end{enumerate}

To prove the first property, consider two
$(\mathcal{V}_a',\mathcal{V}_a'')$ and
$(\mathcal{V}_b',\mathcal{V}_b'')$. Then we can find a row
permutation matrix $\mathbf{E}_{\operatorname{row}}$ and a column
permutation matrix $\mathbf{E}_{\operatorname{col}}$ such that
\begin{equation}
\mathcal{H}^F_{\mathcal{V}_{m_0},\mathcal{V}_{m_0+1}}(\mathbf{G}',\mathcal{V}_a',\mathcal{V}_a'')=\{\mathbf{E}_{\operatorname{row}}\mathbf{H}\mathbf{E}_{\operatorname{col}}\big|\mathbf{H}\in
\mathcal{H}^F_{\mathcal{V}_{m_0},\mathcal{V}_{m_0+1}}(\mathbf{G}',\mathcal{V}_b',\mathcal{V}_b'')\}.
\end{equation}
Therefore, from the fact that
$|\mathcal{V}(\mathbf{H},\mathcal{V}_{m_0},\mathcal{V}_{m_0+1})|=|\mathcal{V}(\mathbf{E}_{\operatorname{row}}\mathbf{H}\mathbf{E}_{\operatorname{col}},\mathcal{V}_{m_0},\mathcal{V}_{m_0+1})|$,
the first property holds.

\begin{figure}[t!]
  \begin{center}
  \scalebox{1.1}{\includegraphics{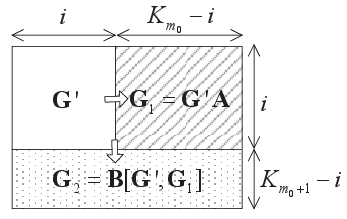}}
  \caption{Construction of $\mathcal{H}^F_{\mathcal{V}_{m_0},\mathcal{V}_{m_0+1}}(\mathbf{G}',\mathcal{V}',\mathcal{V}'')$, where $\mathbf{A}\in\mathbb{F}_2^{i\times (K_{m_0}-i)}$, and $\mathbf{B}\in\mathbb{F}_2^{(K_{m_0+1}-i)\times i}$.}
  \label{FIG:constructive_way}
  \end{center}
\end{figure}

Now consider the second property. We assume that
$\mathcal{V}'=\{v_{1,m_0},\cdots,v_{i,m_0}\}$ and
$\mathcal{V}''=\{v_{1,m_0+1},\cdots,v_{i,m_0+1}\}$ for the proof,
but the same property can be easily derived for arbitrary
$\mathcal{V}'$ and $\mathcal{V}''$ by using the first property. Fig.
\ref{FIG:constructive_way} illustrates the construction of
$\mathcal{H}^F_{\mathcal{V}_{m_0},\mathcal{V}_{m_0+1}}(\mathbf{G}',\mathcal{V}',\mathcal{V}'')$.
We obtain $i\times (K_{m_0}-i)$ matrix
$\mathbf{G}_1=\mathbf{G}'\mathbf{A}$, where
$\mathbf{A}\in\mathbb{F}_2^{i\times(K_{m_0}- i)}$. Then
$(K_{m_0+1}-i)\times K_{m_0}$ matrix $\mathbf{G}_2$ is obtained by
setting $\mathbf{G}_2=\mathbf{B}[\mathbf{G}',\mathbf{G}_1]$, where
$\mathbf{B}\in \mathbb{F}_2^{(K_{m_0+1}-i)\times i}$. Therefore, we
obtain
\begin{equation}
\mathcal{H}^F_{\mathcal{V}_{m_0},\mathcal{V}_{m_0+1}}(\mathbf{G}',\mathcal{V}',\mathcal{V}'')=\Big\{\left[[\mathbf{G}',\mathbf{G}_1]^T,[\mathbf{G}_2]^T\right]^T\big|\mathbf{A}\in\mathbb{F}_2^{i\times(K_{m_0}-
i)},\mathbf{B}\in \mathbb{F}_2^{(K_{m_0+1}-i)\times i}\Big\}.
\end{equation}
Then, for given $\mathbf{A}$ and $\mathbf{B}$,
$\big|\mathcal{V}\big(\left[[\mathbf{G}',\mathbf{G}_1]^T,[\mathbf{G}_2]^T\right]^T,\mathcal{V}_{m_0},\mathcal{V}_{m_0+1}\big)\big|$
is the same for all $\mathbf{G}'$ having the same rank. Therefore,
the second property holds.

From the above two properties,
$\sum_{\mathbf{H}\in\mathcal{H}^F_{\mathcal{V}_{m_0},\mathcal{V}_{m_0+1}}(\mathbf{G}',\mathcal{V}',\mathcal{V}'')}\frac{1}{|\mathcal{V}(\mathbf{H},\mathcal{V}_{m_0},\mathcal{V}_{m_0+1})|}$
is the same for all $\mathcal{V}'$, $\mathcal{V}''$, and
$\mathbf{G}'$ having the same rank. We also know that
$|\mathcal{V}\left(\rank(\mathbf{G}'),\rank(\mathbf{G}'),\mathcal{V}_{m_0},\mathcal{V}_{m_0+1}\right)|$
is the same for all $\mathbf{G}'$ having the same rank. As a result,
$P_m^{(6)}(\mathbf{G}')$ is the same for all $\mathbf{G}'$ having
the same rank. Thus, from (\ref{EQ:p_G_p_H}) and (\ref{EQ:ph}), we
have
\begin{equation}
P_m^{(6)}(\mathbf{G})\sum_{\mathbf{G}'\in\mathcal{F}_i}1=2^{-K_{m_0+1}K_{m_0}}\sum_{\mathbf{H}\in\mathbb{F}^{K_{m_0+1}\times
K_{m_0}}_2,\rank(\mathbf{H})=i}1.
\end{equation}
Since $\sum_{\mathbf{G}'\in\mathcal{F}_i}1=N_{i,i}(i)$ and
$\sum_{\mathbf{H}\in\mathbb{F}^{K_{m_0+1}\times
K_{m_0}}_2,\rank(\mathbf{H})=i}1=N_{K_{m_0+1},K_{m_0}}(i)$, we
finally obtain
\begin{equation}
P_m^{(6)}(\mathbf{G})=2^{-K_{m_0+1}K_{m_0}}\frac{N_{K_{m_0+1},
K_{m_0}}(i)}{N_{i,i}(i)}.
\end{equation}
In conclusion, Lemma \ref{LEM:mapping1}.(2) holds.
\hfill$\blacksquare$

\emph{{~~}Proof of Lemma \ref{LEM:mapping1}.(3):} From the
definitions of $P_m^{(6)}(\mathbf{G})$ and
$P_m^{(9)}(\mathbf{G},\mathcal{V}',\mathcal{V}'')$, we obtain
\begin{eqnarray}
P_m^{(6)}(\mathbf{G})\!\!\!\!\!\!\!\!\!&&=\sum_{(\mathcal{V}',\mathcal{V}'')\in\mathcal{V}(i,i,\mathcal{V}_m,\mathcal{V}_{m+1})}P_m^{(9)}(\mathbf{G},\mathcal{V}',\mathcal{V}'')\nonumber\\
&&={K_m \choose i}{K_{m+1} \choose
i}P_m^{(9)}(\mathbf{G},\mathcal{V}_m',\mathcal{V}_{m+1}'),
\end{eqnarray}
where the second equality holds since
$|\mathcal{V}(i,i,\mathcal{V}_m,\mathcal{V}_{m+1})|={K_m
\choose i}{K_{m+1} \choose i}$ and
$P_m^{(9)}(\mathbf{G},\mathcal{V}',\mathcal{V}'')$ is the same for
all $\mathcal{V}'$ and $\mathcal{V}''$. Thus, we have
\begin{equation}
P_m^{(9)}(\mathbf{G},\mathcal{V}_m',\mathcal{V}_{m+1}')=P_m^{(6)}(\mathbf{G})/\left({K_m
\choose i}{K_{m+1} \choose
i}\right),
\end{equation}
which completes the proof. \hfill$\blacksquare$

\end{document}